\documentclass[twocolumn,amsmath,amssymb,floats,superscriptaddress,nofootinbib,10pt]{revtex4-1}

\pdfoutput=1

\AtBeginDocument{%
    \newwrite\bibnotes
    \def\bibnotesext{Notes.bib}
    \immediate\openout\bibnotes=\jobname\bibnotesext
    \immediate\write\bibnotes{@CONTROL{%
    apsrev41Control,author="08",editor="1",pages="1",title="0",year="1"}}
     \if@filesw
     \immediate\write\@auxout{\string\citation{apsrev41Control}}%
    \fi
}%

\usepackage{graphicx}
\usepackage{dcolumn}
\usepackage{bm}
\usepackage{revsymb}
\usepackage[usenames]{color}
\usepackage{color}
\usepackage{physics}
\usepackage{amsfonts}
\usepackage[section]{placeins} 

\usepackage{tikz}

\usepackage{dblfloatfix} 

\newcommand{\vct}[1]{\mathbf{#1}}

\usepackage{amsmath}
\usepackage{ulem}
\usepackage[usenames]{color}
\usepackage{amsfonts}
\usepackage{graphicx}
\usepackage{dcolumn}
\usepackage{bm}
\usepackage{revsymb}
\usepackage[usenames]{color}
\usepackage{graphicx}

\usepackage{color}
\usepackage{physics}
\usepackage{amsmath}
\usepackage{amssymb,bbm}
\usepackage{lipsum}  
\usepackage{float}

\usepackage{subcaption}

\usepackage[version=4]{mhchem}

\usepackage{xcolor}

\newcommand{\Comment}[3]{\par\noindent\textcolor{#1}{\llap{\footnotesize #2 }\fbox{\parbox{0.98\linewidth}{\textsf{\footnotesize #3}}}}\par}

\newcommand{\bea}{\begin{eqnarray}}
\newcommand{\eea}{\end{eqnarray}}

\newcommand{\kb} {{k_\text{b}}}

\newcommand{\odd} {{\text{o}}}
\newcommand{\even} {{\text{e}}}

\usepackage{bbm}

\definecolor{nblue}{RGB}{28,130,185}
\definecolor{cgreen}{RGB}{76,153,0}
\definecolor{myorange}{RGB}{245,156,74}

\usepackage{hyperref}
\hypersetup{
  colorlinks=true,
  citecolor=magenta,
  urlcolor=-myorange
}

\usepackage{xcolor}
\definecolor{ogreen} {RGB}{71,191,145}
\newcommand{\RL}[1]{\Comment{ogreen}{RL}{#1}} 

\usepackage{mathbbol}

\usepackage{hyperref}
\hypersetup{
  colorlinks=true,
  citecolor=magenta,
  urlcolor=-myorange
}

\newcommand{\quotes}[1]{``#1''}

\begin{document}

\title{Chapman-Enskog expansion for chirally colliding disks}

\author{Ruben Lier}
\email{r.lier@uva.nl}
\affiliation{Institute for Theoretical Physics, University of Amsterdam, 1090 GL Amsterdam, The Netherlands}
\affiliation{Dutch Institute for Emergent Phenomena (DIEP), University of Amsterdam, 1090 GL Amsterdam, The Netherlands}
\affiliation{Institute for Advanced Study, University of Amsterdam, Oude Turfmarkt 147, 1012 GC Amsterdam, The Netherlands}
\author{Paweł Matus}
\affiliation{School of Physics and Astronomy, Tel Aviv University, Tel Aviv, Israel}

\begin{abstract}
\textcolor{black}{
We study a two-dimensional gas of hard disks undergoing chiral, energy- and momentum-conserving collisions. We propose a simple chiral collision rule and show that it can effectively capture collisions of spinning noncircular particles without retaining spin or orientation as kinetic variables. Despite the microscopic breaking of time-reversal symmetry, the \(H\)-theorem is obeyed, guaranteeing relaxation towards equilibrium in the absence of external forces. A Chapman--Enskog expansion yields analytical expressions for the shear and odd viscosity and the thermal conductivity. Theoretical predictions are confirmed by nonequilibrium molecular dynamics simulations.
}
\end{abstract}

\maketitle

\section{Introduction}
The kinetic theory of gases was built on the image of colliding billiard balls, identical and symmetric, their interactions governed only by geometry, conservation laws, and discrete symmetries.
Although these microscopic encounters are reversible, an \textit{arrow of time} emerges at macroscopic scales, breaking time-reversal symmetry. This breaking gives rise to dissipative transport such as shear viscosity, thermal conductivity, and the diffusivity of dyes.

In addition to time-reversal symmetry, another important discrete symmetry in kinetic theory is parity. In two dimensions, it is possible for parity and time-reversal symmetry to be simultaneously broken at the microscopic level, which happens when the system is \textit{chiral}. One of the key transport implications from chirality in two dimensions is a nonvanishing \textit{odd viscosity}, a mysterious transport coefficient that in two dimensions only affects the flow under specific circumstances \cite{Hoyos_2012,delacretaz2017transport,ganeshan2017odd,avron1998odd,PhysRevFluids.9.094101,PhysRevE.108.L023101,PhysRevFluids.9.094101,hosaka2021hydrodynamic, hosaka2021nonreciprocal,10.1063/5.0249623,Lucas_2014,abanov2018odd,mycp-62h3,PhysRevB.108.165429}. In addition, chirality causes \quotes{odd thermal conductivity}, which is also called the Righi-Leduc effect \cite{Leduc1888}. For charged particles, there can be an odd charge conductivity, also called the Hall conductivity \cite{Hall1879}. Chirality typically enters kinetic theory in one of the following two ways:
\begin{enumerate}
    \item Chirality is introduced by turning on a Lorentz force or Coriolis force felt by all the particles~\cite{pellegrino2017nonlocal,berdyugin2019measuring,PhysRevLett.117.166601,levay1995berry,avron1995viscosity,hoyosreview} (see Refs.~\cite{chapman1990mathematical,matus2024nonrelativistictransportframeindifferentkinetic,Pitaevskii1981-xx} for kinetic theory treatments).  
    \item \textcolor{black}{Chirality is introduced through an externally imposed rotation of all particles} \cite{HULSMAN197053,Korving1966,soni2019odd,doi:10.1073/pnas.2219385121,hargus2020time,hargus2021odd,han2021fluctuating,PhysRevLett.130.158201,matus2024molecular,banerjee2017odd} (see Refs.~\cite{KaganMaksimov1962,KNAAP1967643,MORAAL1969455,eren2025collisionalmodeloddfluids} for kinetic theory treatments). 
\end{enumerate} 
\begin{figure}
    \centering
\begin{tikzpicture}
  \node[anchor=south west,inner sep=0] (g) at (0,0)
    {\includegraphics[width=1.0\linewidth]{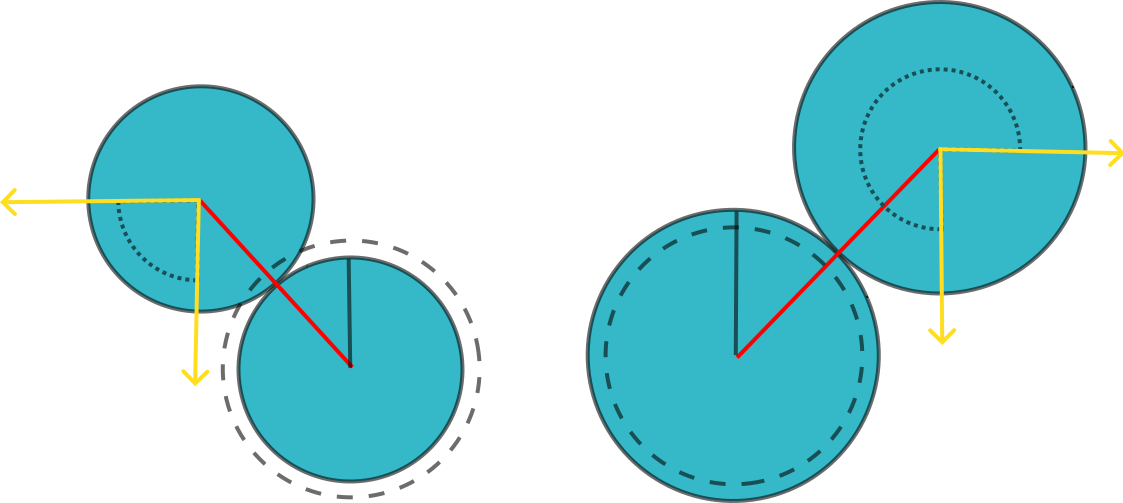}};
  \begin{scope}[x={(g.south east)},y={(g.north west)}]
    \node[fill=none] at (0.82,0.5) {$\mathbf{g} $};
        \node[fill=none] at (0.94,0.76) {$\mathbf{g}' $};
    \node[fill=none] at (0.16,0.32) {$\mathbf{g} $};
        \node[fill=none] at (0.11,0.66) {$\mathbf{g}' $};
        \node[fill=none] at (0.15,0.55) {$ \chi  $};
        \node[fill=none] at (0.23,0.55) {$ \mathbf{n} $};
    \node[fill=none] at (0.75,0.6) {$ \mathbf{n}   $};
        \node[fill=none] at (0.82,0.75) {$ \chi    $};
          \node[fill=white, draw] at (0.40,0.38) {$  r (1  - \varepsilon)  $};
    \node[fill=white, draw] at (0.57,0.45) {$ r (1 + \varepsilon)  $};
            \node[fill=white] at (0.5,0.05) {(b)};
                 \node[fill=white] at (0.10,0.05) {(a)};
  \end{scope}
\end{tikzpicture}
    \caption{Depiction of collision with chiral effective particle radii $r (1 \pm \varepsilon)$. $\mathbf{g}$ is the relative velocity, $\mathbf{n} $ is the vector connecting the centers of the colliding objects, $\mathbf{g}'$ is the outgoing relative velocity, and $\chi$ is the scattering angle. }
    \label{fig:placeholder}
\end{figure}
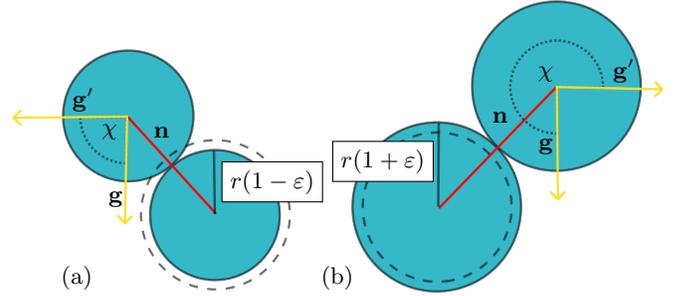
\textcolor{black}{
Systems belonging to the second class are typically modeled by a kinetic theory that couples the translational and rotational degrees of freedom, with the latter coupled to an external field, making analytical calculations difficult. However, because particle spin is not conserved in collisions, it does not correspond to an independent hydrodynamic variable. This points to the possibility that the same phenomenology can be captured by a simpler model of particles colliding in a chiral manner, with the microscopic degrees of freedom effectively integrated out.
\newline 
In this work, we consider a gas of hard disks
\cite{gasssssss,10.1063/1.1761922,PhysRev.127.359}
in which chirality enters intrinsically through the collision criterion:
particles discriminate between left- and right-handed encounters by
biasing whether or not a collision takes place. Crucially, this modifies
only the differential cross section for left- versus right-handed
encounters, while leaving the hard-disk collision rule itself unchanged. 
\newline
This formulation is useful because it retains the main advantages of the
ordinary hard-disk gas. Whenever a collision occurs, it is an ordinary
elastic hard-disk collision, so that energy, momentum, and angular
momentum are conserved. At the same time, the left-right asymmetry of the
collision criterion violates parity and time-reversal symmetry and allows
for odd transport coefficients. We show that, despite this chiral bias, the
\(H\)-theorem is upheld. Hence the model possesses a well-defined
equilibrium state, and transport can be computed by expanding around this
state. In the dilute limit this can be done with the Chapman--Enskog
expansion \cite{chapman1990mathematical}, leading to analytical
predictions for the first-order transport coefficients. The hard-disk
nature of the model also makes it ideally suited for nonequilibrium
molecular dynamics simulations, which we use to corroborate the
Chapman--Enskog results.
}
\section{The model}
\label{sec:model}
Let us now describe the model of chiral hard disks in full detail. We consider a gas of particles of mass $m$, whose collisions conserve energy and momentum. That is, when two particles with ingoing velocity $\mathbf{v}_1$ and $\mathbf{v}_2$ collide, the outgoing velocities $\mathbf{v}'_1$ and $\mathbf{v}'_2$ obey
\begin{subequations} \label{eq:conservation}
\begin{align} \label{eq:moment}
 \mathbf{v}_1 + \mathbf{v}_2    & =  \mathbf{v}'_1 + \mathbf{v}'_2 \,,  \\ 
  v^2_1 + v^2_2   &   =  v^{\prime 2}_1 + v^{\prime 2}_2 \,.  \label{eq:energetic}
 \end{align}
 \end{subequations}
After satisfying \eqref{eq:conservation}, the only thing that is left free is the orientation of the outgoing relative velocity with respect to the ingoing one, i.e. 
 \begin{align}  \label{eq:rottttt}
    \mathbf{g}' = \mathbf{R}( \chi ) \cdot    \mathbf{g} \,, 
 \end{align}
 where $\vct g=\vct v_2-\vct v_1$ and $\vct g'=\vct v'_2-\vct v'_1$ are the relative velocities and $\mathbf{R}( \chi )$ is the two-dimensional rotation matrix that depends only on the scattering angle $\chi$. Hard disk collisions occur instantly when two particles start to overlap. For perfectly smooth hard disks, the forces are exerted along the line of disk centers $\vct n$, and consequently angular momentum is conserved as well.
 
 To introduce chirality without abandoning the simplicity of the hard disk model, we consider the following toy model. Based on the chirality of the collision $c$, which is given by
 \begin{align}
     c = \text{sign} \left[ \hat{\mathbf{z}} \cdot    ( \mathbf{n} \cross \mathbf{g} )  \right], 
     \label{eq:chirality}
 \end{align}
we attribute a different \textit{effective radius} $r(1+c\epsilon)$ to colliding particles, see Fig.~\ref{fig:placeholder}. Considering the geometry of a hard disk collision, one finds that the impact parameter $b$ and the scattering angle $\chi$ are related through
\begin{equation}
\begin{split}
 b &= d (1 + \varepsilon) \cos(\chi/2 ),\qquad     b > 0 ,  \\
b &=  d  (1   -  \varepsilon) \cos(\chi/2 ), \qquad b < 0, 
\end{split} \label{eq:impact}
\end{equation} 
where $d=2r$ is the average diameter. Because the chirality only affects when the particles collide but not the collision rule itself, the model automatically upholds all conservation laws.

The concept of the chirality $c$ of a collision was used in Ref.~\cite{Zhao2022} to model ratchets in a dense suspension. There, the ratchets were modelled as hard disks that are perfectly smooth or infinitely rough depending on chirality $c$, which allowed the authors of Ref.~\cite{Zhao2022} to numerically obtain a nonvanishing odd viscosity. Here, in order to illustrate how our toy model compares with a particular realization of a chirally colliding gas, we consider collisions of rotating ratchet-shaped particles such as shown in Fig.~\ref{fig:placeholder123123}. \textcolor{black}{We expect this setup to capture the generic properties of systems where chiral transport emerges through a combination of a noncircular particle shape and a nonzero average rotation \cite{HULSMAN197053,Korving1966,soni2019odd,doi:10.1073/pnas.2219385121,hargus2020time,hargus2021odd,han2021fluctuating,PhysRevLett.130.158201,matus2024molecular,banerjee2017odd,KaganMaksimov1962,KNAAP1967643,MORAAL1969455}.} As described in the Supplementary Material, we perform a numerical collision experiment, where we consider an ensemble of particles with velocities $\vct v$ and angular velocities $\omega$ which follow the biased Maxwell-Boltzmann distribution 
 \begin{align}
    f_{\mathrm{MB}} (\omega ,  \mathbf{v} ) \sim \exp\left(  -  \frac{ m  \vct v^2  +  I   ( \omega - \Omega  )^2   }{  2  \kb T  }\right), 
    \label{eq:MB_distr}
\end{align}
where $m$ is the particle's mass, $I$ is the moment of inertia, $T$ is the temperature, and $\Omega$ is the average rotation frequency. \textcolor{black}{To account for the fact that the collision rate of particles for a given relative velocity is proportional to the relative velocity magnitude, the corresponding weight factor is added to the sampler.}
We let these particles collide at all impact parameters, keeping track of the scattering angle $\chi $. The corresponding histogram of scattering angles is shown in Fig.~\ref{fig:placeholder2}\textcolor{black}{, where we compare it with the differential cross section of our chiral hard disk model. We observe that our model roughly matches the results except for scattering angles close to $\chi\approx0^\circ$, where the discontinuity due to the abrupt transition of the chiral effective radius is effectively smoothened out.} 
 \begin{figure}
    \centering
\begin{tikzpicture}
  \node[anchor=south west,inner sep=0] (g) at (0,0)
    {\includegraphics[width=0.8\linewidth]{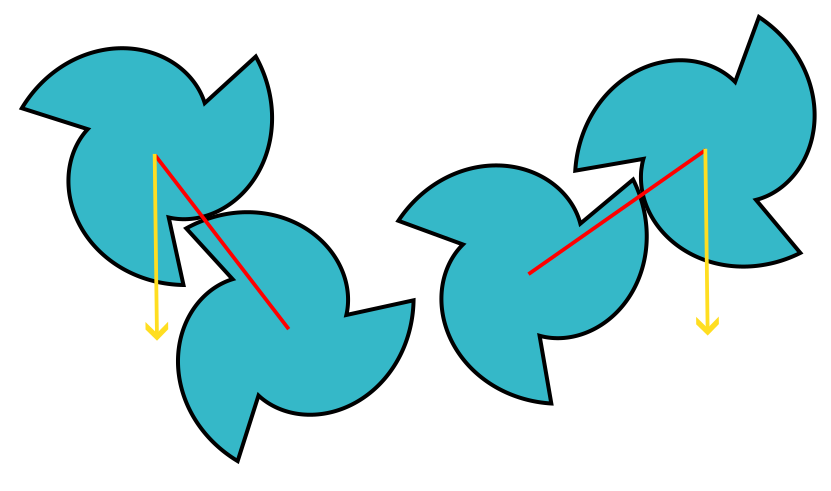}};
  \begin{scope}[x={(g.south east)},y={(g.north west)}]
    \node[fill=none] at (0.14+0.73,0.55) {$\mathbf{g} $};
    \node[fill=none] at (0.15,0.6) {$\mathbf{g} $};
        \node[fill=none] at (0.79-0.48,0.5) {$ \mathbf{n} $};
    \node[fill=none] at (0.7,0.45) {$ \mathbf{n}   $};
            \node[fill=white] at (0.55,0.1) {(b)};
                 \node[fill=white] at (0.15,0.1) {(a)};
  \end{scope}
\end{tikzpicture}
    \caption{Schematic picture of two ratchet-shaped particles moving towards each other in two chirally distinct ways. }
    \label{fig:placeholder123123}
\end{figure}
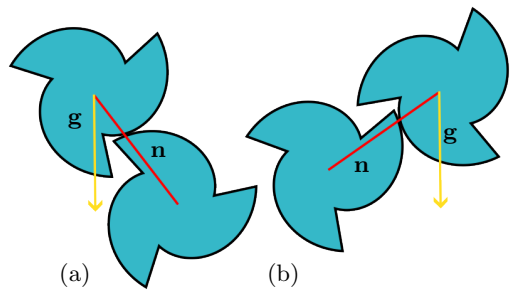
\begin{figure}
    \centering
\begin{tikzpicture}
  \node[anchor=south west,inner sep=0] (g) at (0,0)
{\includegraphics[width=0.9\linewidth]{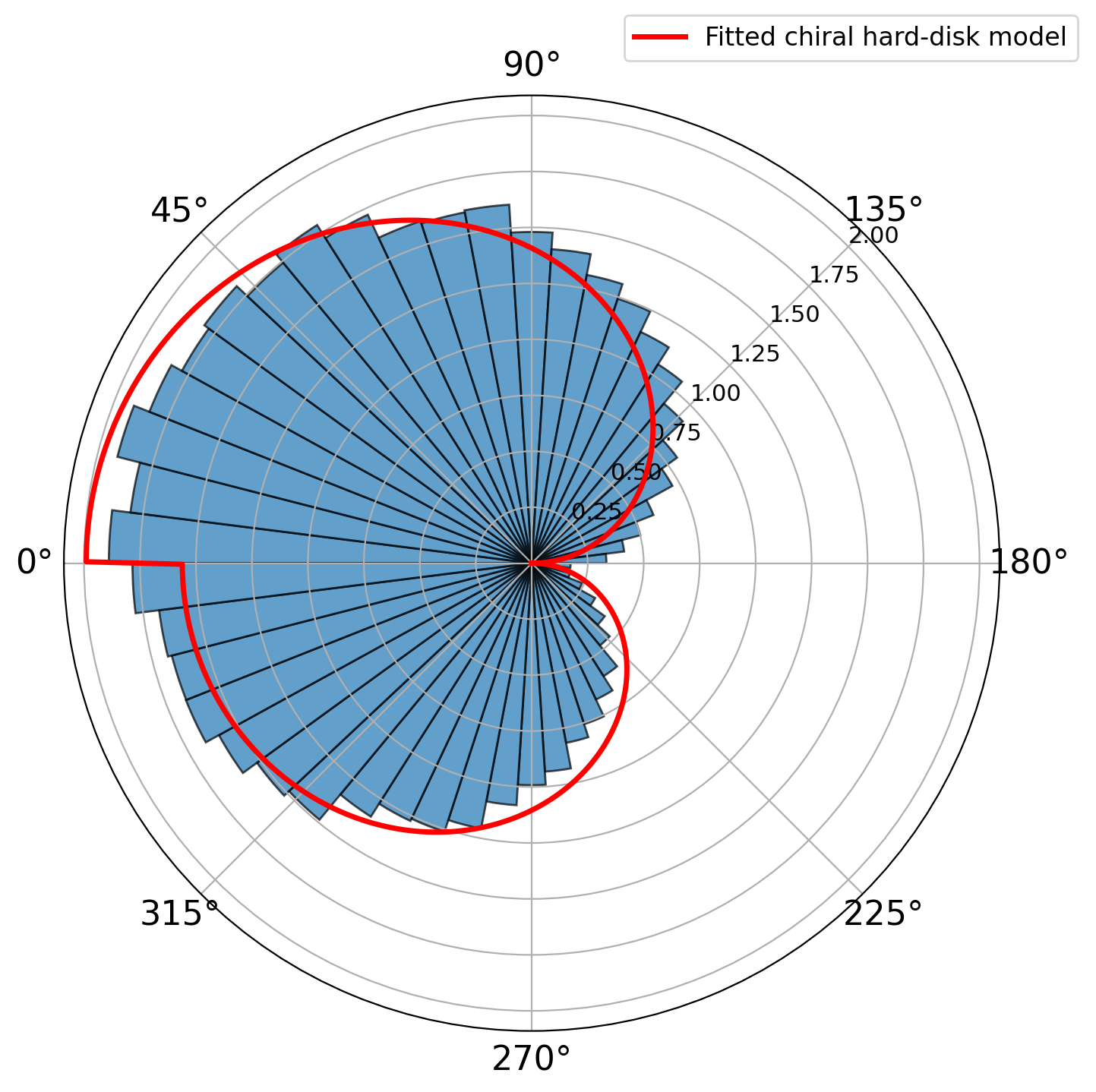}};
  \begin{scope}[x={(g.south east)},y={(g.north west)}]
        \node[fill=none] at (0.875,0.935) {$  $};
  \end{scope}
\end{tikzpicture}
    \caption{Differential scattering cross section as a function of scattering angle $\chi$ corresponding to the numerical collision experiment of ratchet-shaped particles (see Fig.~\ref{fig:placeholder123123}). The particles follow the biased Maxwell-Boltzmann distribution of Eq.~(\ref{eq:MB_distr}) with \textcolor{black}{$\Omega = 0.4\sqrt{2 \kb  T/I}$}. The red line corresponds to the corresponding differential scattering cross section for the collision rule with chiral effective radii (see Fig.~\ref{fig:placeholder}), \textcolor{black}{where we fitted $r=1.78$ and $\varepsilon = 0.12$.} For more details see the Supplementary Material.}
    \label{fig:placeholder2}
\end{figure}

Having specified the nature of the chiral collisions, we can formulate the Boltzmann equation which describes the evolution of the phase-space distribution $f(\mathbf{v},\mathbf{x},t )$ of the disks in the dilute limit. The Boltzmann equation is given by
\begin{align}  \label{eq:fequation}
\begin{split}
 &        \mathcal{L}[f(\mathbf{v_1 })] =        C [f(\mathbf{v_1 })] . \end{split}
\end{align}
The left-hand side of the Boltzmann equation~(\ref{eq:fequation}) is called the streaming term, while the right-hand side is called the collision term. For the type of hard disks considered in this work, these terms are given by 
\begin{subequations}
    \begin{align}
   &   \mathcal{L}[f(\mathbf{v_1 })]    =   \partial_t  f (\mathbf{v}_1)   + \mathbf{v}_1 \cdot \nabla f  (\mathbf{v}_1) ~~  , \\ 
     \begin{split}
              &   C [f(\mathbf{v_1 })]   =  \\ &    \int d \mathbf{v}_2     \int_{ -  d (1   - \varepsilon  ) }^{   d (1  + \varepsilon  )  } d b   |\vct g |  \left(f  (\mathbf{v}_1') f (\mathbf{v}'_2) -f  (\mathbf{v}_1) f (\mathbf{v}_2)\right)   ,   \label{eq:collisionion}
                     \end{split}
\end{align}
\end{subequations}  
where we omitted the dependence of the phase-space distribution on $t$ and $\mathbf{x}$ for simplicity. Following Eq.~(\ref{eq:impact}), Eq.~\eqref{eq:collisionion} can be written as
\begin{align}
\begin{split}
 &     C[f(\mathbf{v}_1)]=  \int d \mathbf{v}_2  \sum_{ \pm}    \frac{d  (1 \pm \varepsilon ) }{2}  \int_{ \frac{\pi}{2}   \mp  \frac{\pi}{2} }^{\frac{3 \pi}{2}   \mp \frac{\pi}{2} } d \chi    \\ 
 & \cdot \sin(\chi / 2 ) |\vct g  | \left(f  (\mathbf{v}_1') f (\mathbf{v}'_2) -f  (\mathbf{v}_1) f (\mathbf{v}_2)\right)  . \end{split} \label{eq:full_boltzmann}
\end{align}
\section{Chapman-Enskog expansion}
Due to their chiral nature, the collisions described in Sec.~\ref{sec:model} break both parity and time reversal -- two symmetries commonly assumed in the derivation of the $H$-theorem \cite{degroot1984nonequilibrium,Cercignani1988}. What is more, the collision rules in our toy model do not follow from any Hamiltonian. Nevertheless, as long as the collisions conserve energy and momentum, the $H$-theorem must hold, as we rigorously prove in the Supplementary Material. Therefore, the system possesses a well-defined thermal equilibrium in which the particles obey the Maxwell-Boltzmann distribution. This observation forms the foundation of the hydrodynamic approach to transport. The system is assumed to be in the vicinity of a \textit{local} equilibrium state described by the distribution function
\begin{equation}
    f_0(\vct v,\vct x, t) = \frac{ m n(\vct x,t) }{2 \pi \kb T (\vct x,t) }  \exp\left(-\frac{m (\vct v - \vct u(x,t))^2}{ 2 \kb  T(\vct x,t)}\right).
    \label{eq:maxwell_distr}
\end{equation}
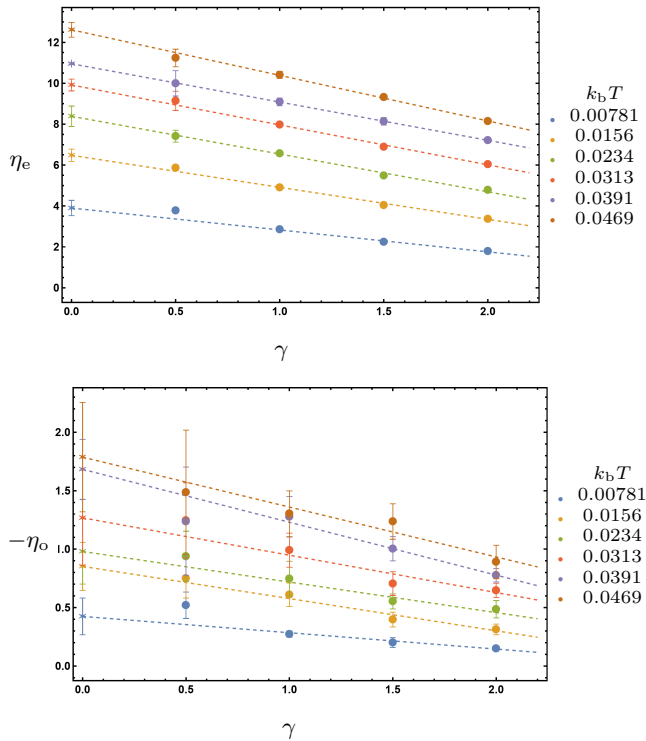
\begin{figure}
\centering
\begin{tikzpicture}
  \node[anchor=south west,inner sep=0] (g) at (0,0)
    {\includegraphics[width=0.8\linewidth]{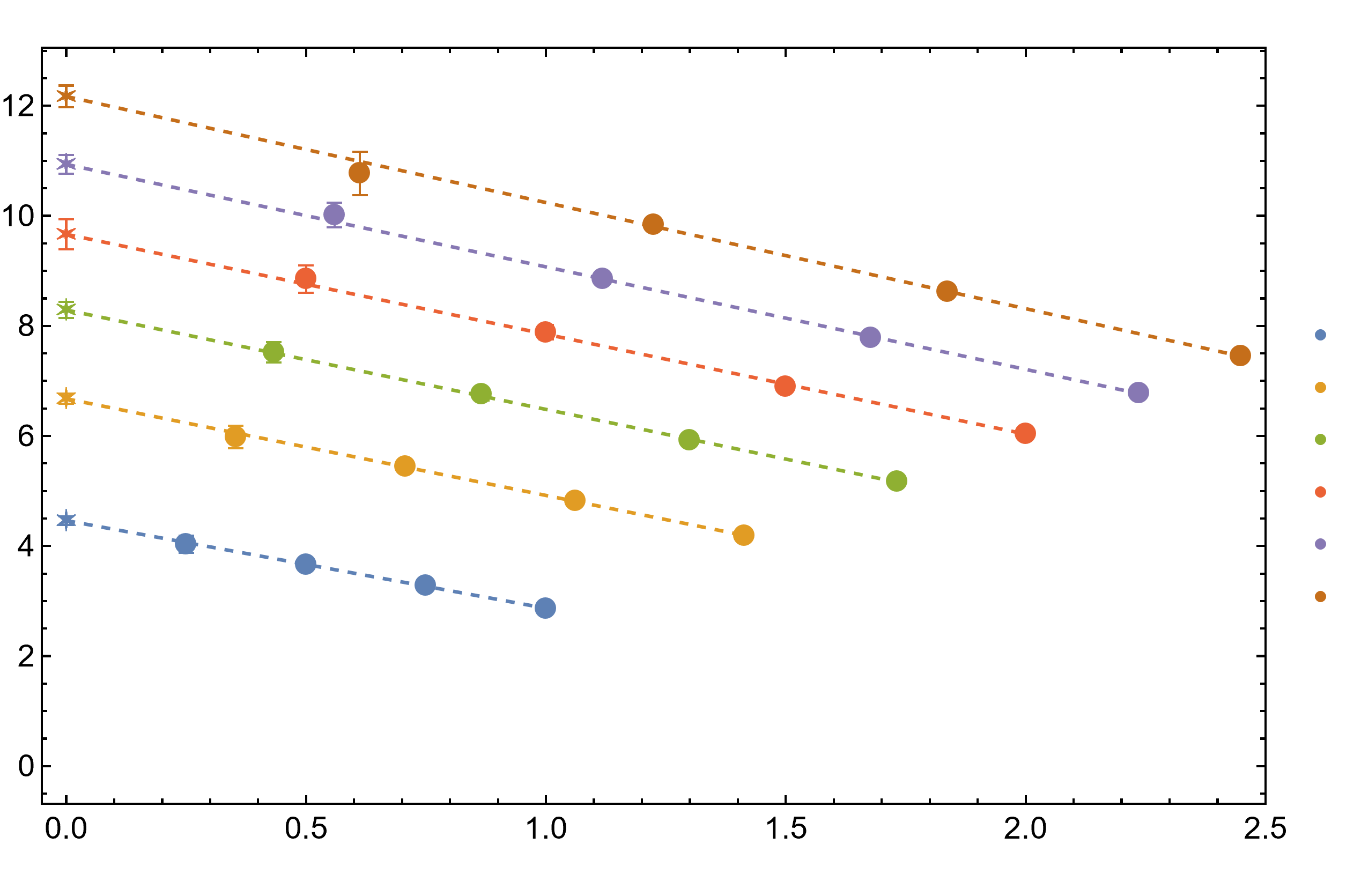}};
  \begin{scope}[x={(g.south east)},y={(g.north west)}]
    \node[fill=white] at (0.45,-0.05) {$\gamma$};
    \node[fill=white] at (-0.05,0.5) {$ \eta_{\even }$};
\node[fill=none] at (1.07,0.70-0) {\scriptsize $\kb T$};
\node[fill=none] at (1.07,0.64-0.0) {\scriptsize $0.00781$};
\node[fill=none] at (1.07,0.58-0.0) {\scriptsize $0.0156$};
\node[fill=none] at (1.07,0.52-0.0) {\scriptsize $0.0234$};
\node[fill=none] at (1.07,0.46-0.0) {\scriptsize $0.0313$};
\node[fill=none] at (1.07,0.40-0.0) {\scriptsize $0.0391$};
\node[fill=none] at (1.07,0.34-0.0) {\scriptsize $0.0469$};
  \end{scope}
\end{tikzpicture}
\vspace{1em}
\begin{tikzpicture}
  \node[anchor=south west,inner sep=0] (g) at (0,0)
    {\includegraphics[width=0.8\linewidth]{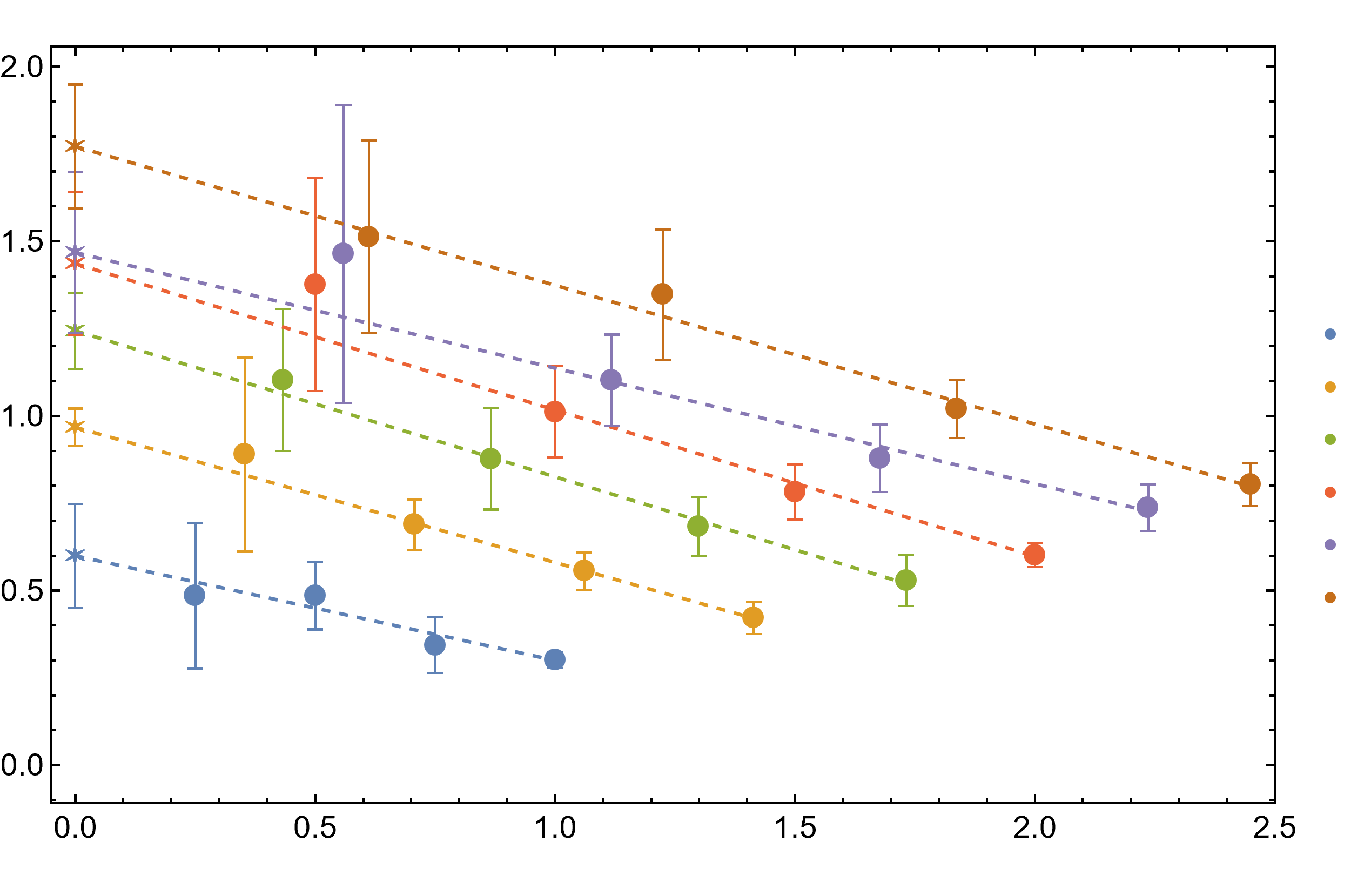}};
  \begin{scope}[x={(g.south east)},y={(g.north west)}]
    \node[fill=white] at (0.45,-0.05) {$\gamma$};
\node[fill=white] at (-0.05,0.5) {$ - \eta_{\odd}$};

\node[fill=none] at (1.07,0.70-0) {\scriptsize $\kb T$};
\node[fill=none] at (1.07,0.64-0.0) {\scriptsize $0.00781$};
\node[fill=none] at (1.07,0.58-0.0) {\scriptsize $0.0156$};
\node[fill=none] at (1.07,0.52-0.0) {\scriptsize $0.0234$};
\node[fill=none] at (1.07,0.46-0.0) {\scriptsize $0.0313$};
\node[fill=none] at (1.07,0.40-0.0) {\scriptsize $0.0391$};
\node[fill=none] at (1.07,0.34-0.0) {\scriptsize $0.0469$};
  \end{scope}
\end{tikzpicture}
\caption{Extrapolation of numerically obtained shear viscosity $\eta_{\even}$ (top) and odd viscosity $\eta_{\odd}$ (bottom) to infinitesimal shear rate $\gamma$ for $\varepsilon =0.5$.}
\label{fig:result1}
\end{figure} 
Here, velocity $\vct u(\vct x,t)$, disk density $n(\vct x,t)$, and temperature $T(\vct x,t)$ are functions of time and space and are assumed to vary slowly in comparison to the typical relaxation time and the mean free path for collisions. Importantly, any distribution function of the form~(\ref{eq:maxwell_distr}) is a solution of the collisional part of the Boltzmann equation: $\mathcal{C}[f_0]=0$, but the corresponding streaming term is generically nonzero: $\mathcal{L}[f_0]\neq 0$. The Chapman-Enskog expansion assumes that the actual solution to the Boltzmann equation can be approximated by a series of perturbations to $f_0(\vct v,\vct x,t)$:
\begin{equation}
    f(\vct v, \vct x, t) = f_0(\vct v,\vct x,t)\left[1+\Phi(\vct v, \vct x,t)+\cdots\right],
\end{equation}
where each successive term is proportional to higher gradients of $\vct u(\vct x,t)$ and $T(\vct x,t)$. The first-order correction $\Phi(\vct v, \vct x,t)$ is found by solving the equation
\begin{equation} \
  \mathcal{L}[f_0] = f_0   C^{(1) } [ \Phi ],
    \label{eq:Phi}
\end{equation}
where the right-hand side is the collision term $\mathcal{C}[f_0(1+\Phi)]$ expanded to the linear order in $\Phi(\vct v,\vct x,t)$:
\begin{widetext}
\begin{align}
\begin{split}
 &         C^{(1)}[\Phi](\vct v_1) =  -\int d \mathbf{v}_2 f_0(\vct v_2)  \sum_{ \pm}    \frac{d  (1 \pm \varepsilon ) }{2}  \int_{ \frac{\pi}{2}   \mp  \frac{\pi}{2} }^{\frac{3 \pi}{2}   \mp \frac{\pi}{2} } d \chi    \sin(\chi / 2 ) |\vct g  | \left[\Phi(\vct v_1)+\Phi(\vct v_2)-\Phi(\vct v_1')-\Phi(\vct v_2')\right]  . \end{split} \label{eq:coll_operator}
\end{align}
\end{widetext}
In the equation above we dropped the arguments $\vct x$ and $t$ since they are equal for all the functions. Calculating the transport coefficients from \eqref{eq:Phi} using the Chapman-Enskog theory involves a truncated expansion of $\Phi$ in Sonine polynomials, which is worked out in the Supplementary Material. We find the following viscosities $\eta_{\even , \odd}$ and thermal conductivities $\kappa_{\even , \odd}$ for a Sonine expansion truncated at order $N$:
\begin{equation}  \label{eq:finalresult2}
\begin{split}
    \eta^{(N)}_\even & =   \eta^{(0)}_\even N^{(N)}_\even (\varepsilon) , ~~ \eta^{(N)}_\odd = \eta^{(0)}_\odd N^{(N)}_\odd (\varepsilon) , \\
    \kappa^{(N)}_\even & = \kappa^{(0)}_\even K^{(N)}_\even (\varepsilon) , ~~  \kappa^{(N)}_\odd = \kappa^{(0)}_\odd  K^{(N)}_\odd (\varepsilon)  , 
\end{split}
\end{equation}
where the transport coefficients at zeroth order in a Sonine polynomial expansion are given by
\begin{equation}  \label{eq:finalresult}
\begin{split}
    \eta^{(0)}_\even & = \frac{8}{d(16+\varepsilon^2)}\sqrt{\frac{m \kb T}{\pi  }} , ~~ \eta^{(0)}_\odd = -\frac{2\varepsilon}{d(16+\varepsilon^2)}\sqrt{\frac{m \kb T}{\pi  }}  , \\
    \kappa^{(0)}_\even & = \frac{32 \kb T }{ d(16+\varepsilon^2)}\sqrt{\frac{\kb T}{\pi m }}  , ~~  \kappa^{(0)}_\odd = -\frac{8\varepsilon \kb T }{ d(16+\varepsilon^2)}\sqrt{\frac{\kb T}{\pi m }}, 
\end{split}
\end{equation}
whereas the functions $N^{(N)}_{\even , \odd } (\varepsilon)$ and $K^{(N)}_{\even , \odd } (\varepsilon)$ are multiplicative corrections corresponding to the order at which the Sonine polynomial expansion is truncated. Taking $N=5$ and $\varepsilon  =  0.5$, we find
\begin{equation}
\begin{split}
N^{(5)}_\even (0.5)   & = 1.02196      , ~~  N^{(5)}_\odd (0.5)  =   1.02166  , \\
K^{(5)}_\even (0.5)   &  =    1.02994    , ~~  K^{(5)}_\odd (0.5)   =  1.02002   . 
\end{split} \label{eq:sonine04}
\end{equation}
Since these multiplicative correction coefficients are close to one, the truncation error can be assumed to be small.
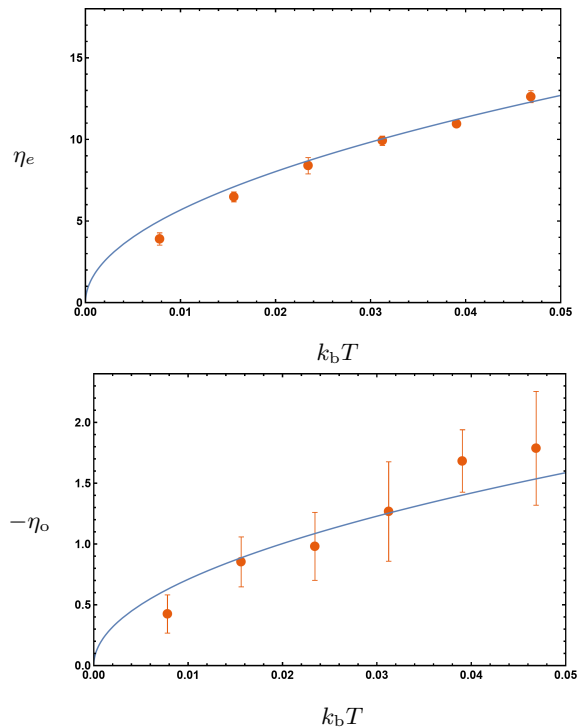
\begin{figure}
\centering
\begin{tikzpicture}
  \node[anchor=south west,inner sep=0] (g) at (0,0)
    {\includegraphics[width=0.9\linewidth]{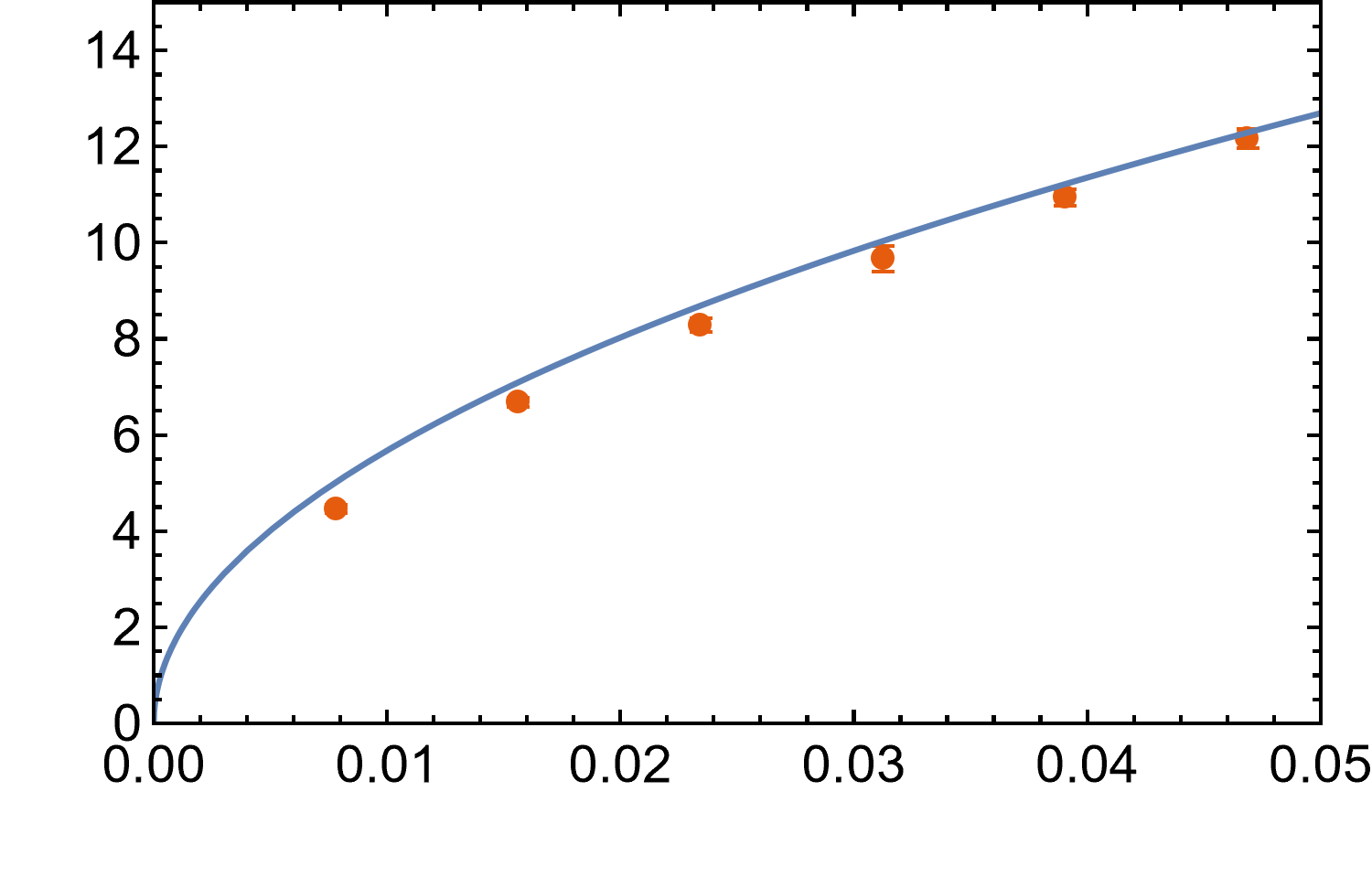}};
  \begin{scope}[x={(g.south east)},y={(g.north west)}]
    \node[fill=white] at (0.55,-0) {$\kb T $};
    \node[fill=white] at (-0.0,0.54) {$\eta_{e} $};
  \end{scope}
\end{tikzpicture}
\vspace{1em}
\centering
\begin{tikzpicture}
  \node[anchor=south west,inner sep=0] (g) at (0,0)
    {\includegraphics[width=0.9\linewidth]{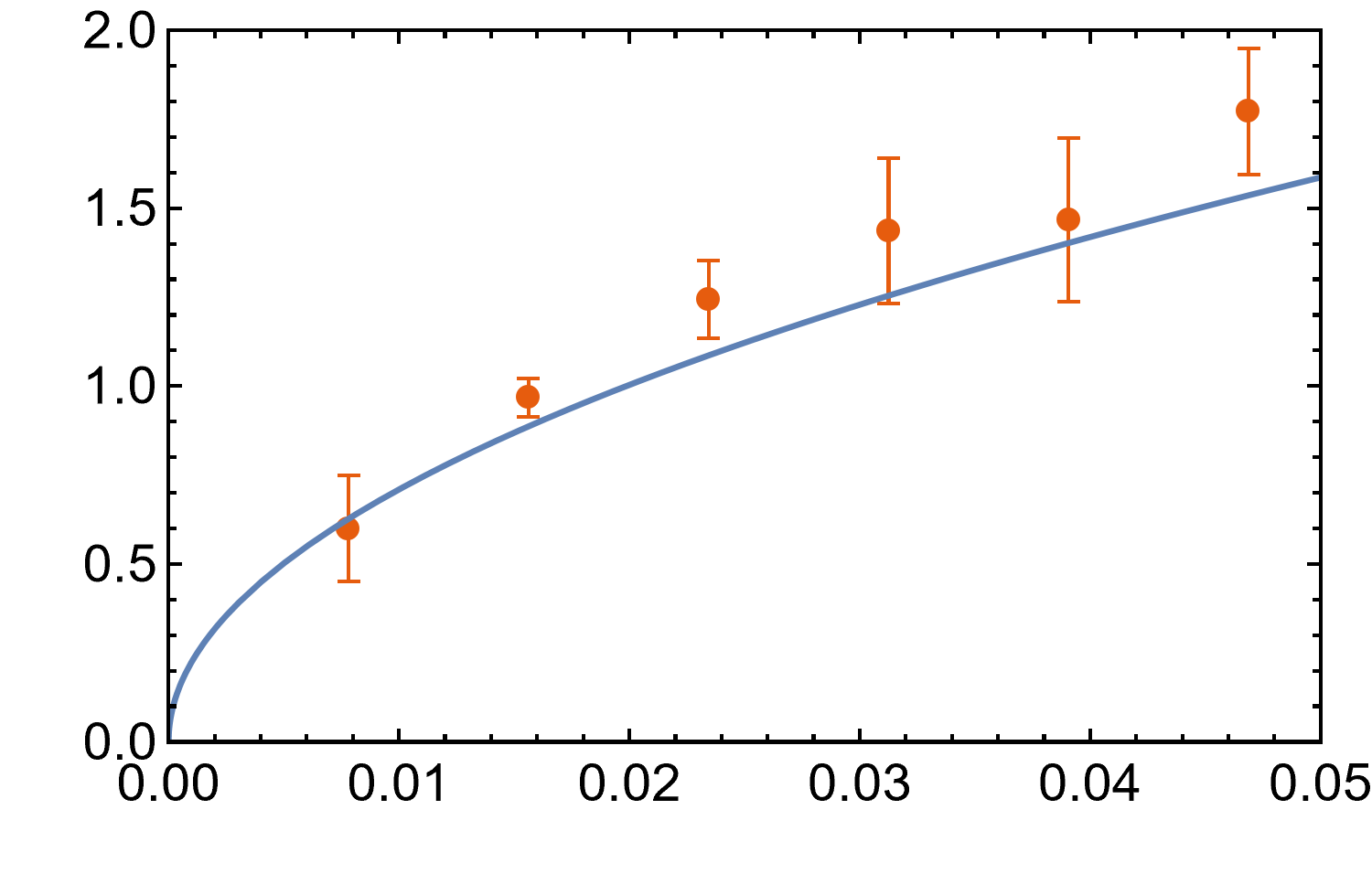}};
  \begin{scope}[x={(g.south east)},y={(g.north west)}]
    \node[fill=white] at (0.55,-0.0) {$\kb T $};
    \node[fill=white] at (-0.0,0.54) {$- \eta_{\mathrm{o}}  $};
  \end{scope}
\end{tikzpicture}
\caption{Shear viscosity $\eta_{\mathrm{e}}$ (top) and odd viscosity $\eta_{\mathrm{o}}$ (bottom) as a function of $\kb T$ for $\varepsilon=0.5$ \textcolor{black}{and packing fraction $\phi=0.02$}. The numerical results were obtained using the extrapolation shown in Fig.~\ref{fig:result1}.}
\label{eq:result2}
\end{figure}

To verify \eqref{eq:finalresult}, we use nonequilibrium molecular dynamics. In particular, we consider a shear periodic box of hard disks which collide with the collision rule \textcolor{black}{of Fig.}~\ref{fig:placeholder}. We turn on shear with the SLLOD algorithm \cite{Evans2007-dg,PhysRevA.30.1528,10.1063/1.2192775,han2021fluctuating}. The details of this simulation are provided in the Supplementary Material and the full code and simulation output is available at \url{https://github.com/rubenlier/chiral-chapman-enskog}. When implementing the SLLOD algorithm, one faces a tradeoff, where increasing the shear rate $\gamma$ on the one hand accelerates convergence of the result, and on the other hand introduces nonlinear corrections related to shear thinning \cite{Evans2007-dg}. For this reason, to get rid of this systematic error, we compute the shear and odd viscosity at different values of shear rate and extrapolate to the regime where shear rate is infinitesimal (see Fig.~\ref{fig:result1}). \textcolor{black}{Note that we chose our shear rate extrapolation range to be $\propto \sqrt{T}$ to account for the fact that, at low temperatures, high shear rate values affect viscosity nonlinearly, whereas at high temperatures, low shear rate values yield noisy viscosities.} The resulting shear and odd viscosities are given in Fig.~\ref{eq:result2}. 

\textcolor{black}{We see in Fig.~\ref{eq:result2} that the overlap between numerical results and the theory is strong especially for shear viscosity, whereas we find for odd viscosity a \textcolor{black}{larger} statistical error. This \textcolor{black}{larger} statistical error is explained by the fact that, as follows from \eqref{eq:finalresult}, the odd viscosity considered in this work is always one order of magnitude smaller than shear viscosity, which strongly lowers the signal-to-noise ratio.} 


\section{Discussion}
In this work, we introduce a new microscopic approach to odd viscosity, which relies neither on a background parity-odd field nor on the external inflow of angular momentum. Instead, the approach is based on hard disks with chiral effective radii. Because the collision rule obeys particle number, energy, and momentum conservation, it is possible to find the equilibrium state, with respect to which we expand to compute parity-even and parity-odd transport. The resulting expressions for viscosity are compared to a computation of viscosity using nonequilibrium molecular dynamics simulations.
\newline 
\textcolor{black}{Outside of our work, other analytically tractable kinetic theory models of chiral transport have recently been developed~\cite{fruchart2022odd,jiao2025hydrodynamicsthreedimensionalmesoscaleodd,gkct-3247}. In particular, a closely related route was taken in Ref.~\cite{fruchart2022odd}, where transport coefficients were derived assuming differential scattering cross section of a hard disk rotated by a constant angle. However, an attempt to apply that model in many-particle simulations would reveal an inherent issue: particles try to move closer to one another after colliding at a grazing angle. Additionally, angular momentum is locally injected into the system at each collision. In contrast, the model in Refs.~\cite{jiao2025hydrodynamicsthreedimensionalmesoscaleodd,gkct-3247} is easily amenable to numerics, but inherently mesoscopic and agnostic to the microscopic source of chirality. To the best of our knowledge, our work is the first one to propose a microscopic model of a chiral collision of classical particles motivated by realistic scattering mechanisms and validate the results in a many-particle simulation.
\newline 
The kinetic theory studied here also suggests a natural experimental setting.
In the air-table experiments of Ref.~\cite{Workamp2018}, a two-dimensional gas
is formed from noncircular disks with chiral air channels, rotated by air flow.
Because momentum loss to the environment is negligible, this system realizes a
gas-like regime and was observed to relax to a Maxwell-Boltzmann velocity
distribution \cite{Workamp2018}. The similarity to the rotating ratchet setup considered here makes it a particularly relevant platform
for testing the transport predictions of our model.
\newline 
In future work, it would be interesting to establish for odd thermal
conductivity the same quantitative connection between analytics and numerics
as was made here for odd viscosity. Another natural direction is odd
self-diffusion
\cite{hargus2021odd,Kalz_2022,PhysRevLett.132.057102,k8vq-dw2w,caprini2025activethermodynamicsinertialchiral,Vega_Reyes_2022,10.1063/5.0095969},
which would require a Boltzmann--Lorentz formulation of the chiral kinetic theory introduced in this work in which tagged particles
collide with an equilibrium bath.} 

\section{Acknowledgements}
We thank Tobias Holder, Tomer Markovich, Naomi Oppenheimer and Jan V. Sengers for useful discussions. \textcolor{black}{We thank the anonymous referee for showing that the present kinetic
framework can be adapted to compute odd self-diffusion of particles in a bath.} R.L. was supported by the NWA ORC programme Emergence at All Scales. P.M. was supported by the European Research Council (ERC) under grant QuantumCUSP (Grant Agreement No. 101077020), the Harry Bloomfield International Scholarship, and a postdoctoral fellowship of the Azrieli foundation.

\appendix

\onecolumngrid

\newpage

\section*{Numerical collision experiment}
\label{eq:numericalcollisionexperiment}

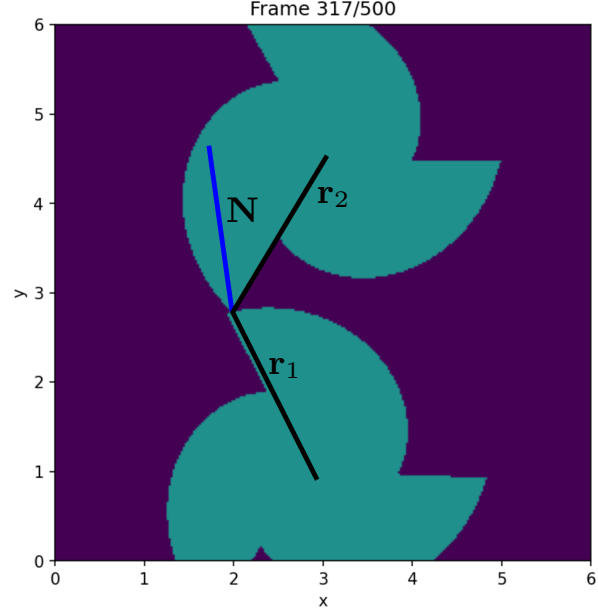
\begin{figure}
       \centering
\begin{tikzpicture}
  \node[anchor=south west,inner sep=0] (g) at (0,0)
{\includegraphics[width=0.45\linewidth]{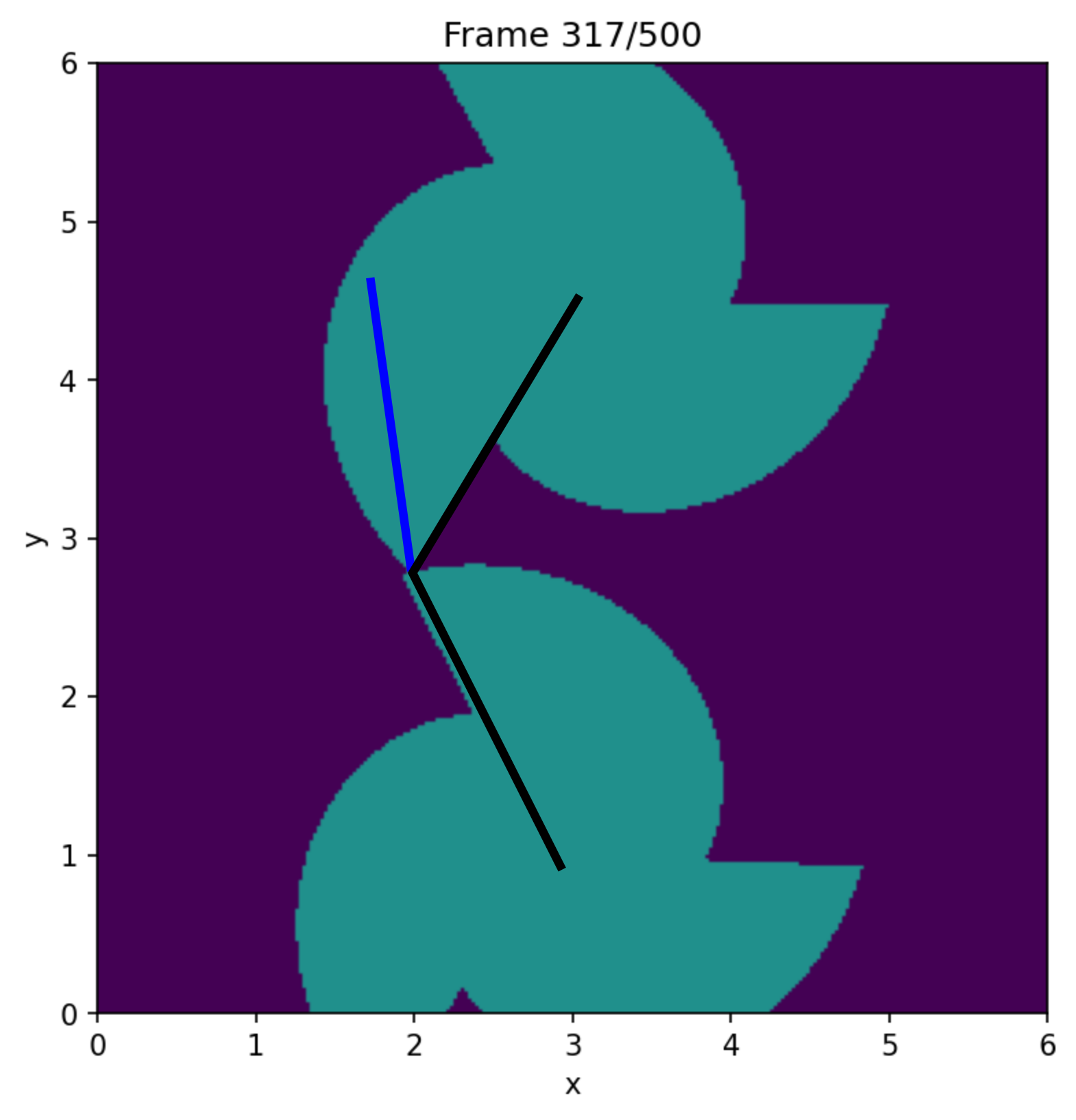}};
  \begin{scope}[x={(g.south east)},y={(g.north west)}]
        \node[fill=none,scale=1.5] at (0.47,0.4) {$\mathbf{r}_1  $};
        \node[fill=none,scale=1.5] at (0.55,0.67) {$\mathbf{r}_2  $};
      \node[fill=none,scale=1.5] at (0.4,0.65) {$\mathbf{N}  $};
  \end{scope}
\end{tikzpicture}
    \caption{Snapshot of ratchet-shaped particles colliding, showing the normal vector $\mathbf{N}$ and the vectors $\mathbf{r}_{1,2}$ that give the distance of the collision point from the respective ratchet centers. Here $R_{\mathrm{inner}}=1.0$, $R_{\mathrm{outer}}=2.0$, $b=3.5$.}
    \label{fig:placeholder1}
\end{figure}

In this section we describe the numerical collision experiment that aims to determine the scattering cross section of ratchet-shaped particles with a nonzero average rotation $\Omega$ and fit to it the parameters $r$ and $\varepsilon$ of the toy model introduced in Sec.~\ref{sec:model}. To do this, we initialize two particles with the sawblade shape presented in Fig.~\ref{fig:placeholder123123}, characterized by the inner and outer radius $R_{\text{inner}}$ and $R_{\text{outer}}$, initial rotation angles $\psi^{(0)}_{1,2}$, and angular velocities $\omega_{1,2}$. In the polar coordinate system centered on a given particle, the boundary radius of that particle is given by the formula
\begin{align}  \label{eq:boundarycoord}
    r_{1,2} (\phi ) =  R_{\text{inner}} + (R_{\text{outer}} - R_{\text{inner}}  ) \cdot   \text{mod} ( 3  \psi_{1,2} (\phi) / 2 \pi  , 1   ) \, , 
\end{align}
where the phases $\psi_{1,2}$ evolve in time as
\begin{align}
    \psi_{1,2} (\phi)  = \phi  -  \psi^{(0)}_{1,2}   -  \omega_{1,2} t.
\end{align} 
Furthermore, the centers of mass of the two particles are displaced by the distance $b$ along the $\vct y$ direction, and they perform uniform motion with velocities $\vct v_{1,2}$ given by $\vct v_{2}=\frac{1}{2} g\hat{\vct{x}} =-\vct v_1$, where $g$ is the magnitude of the relative velocity. For reference, see Fig.~\ref{fig:placeholder1}. A single simulation is initialized with random initial phases $\psi^{(0)}_{1,2}$ drawn from a uniform distribution $\psi^{(0)}_{1,2} \in (0,2\pi)$, an impact parameter $b$ drawn from a uniform distribution $b\in(- 2 R_{\text{outer}},  2 R_{\text{outer}})$, and the relative velocity $g$ and rotation frequencies $\omega_{1,2}$ sampled according to
\begin{align}
   \textcolor{black}{ P (\omega_{1,2} ,  g ) \sim  g^2  \exp\left(  -  \frac{ \frac{1}{2} m  g^2  +  \sigma m R_{\text{outer}}^2   ( \omega_{1,2} - \Omega  )^2   }{  2  \kb T  }\right) \, ,  }  
\end{align}
where $\sigma$ is a dimensionless number such that the particle's moment of inertia is given by
\begin{align}
    I = \sigma m  R_{\text{outer}}^2\, .  
\end{align}
We consider only the case where the particles move towards each other. The evolution of the particles' shapes according to Eq.~(\ref{eq:boundarycoord}) continues until the particles collide (see Figure~\ref{fig:placeholder1}), after which we compute the postcollisional velocities which allow us to extract the scattering angle. In order to compute the postcollisional velocities, we assume that the particles are perfectly smooth, which means that only the normal total relative velocity $\vct v_{\mathrm{rel}}$ of the particle is flipped, keeping the tangential one unchanged, i.e. 
 \begin{subequations}  \label{eq:Nvector}
\begin{align}
 \mathbf{N} \cdot   \vct v_{\mathrm{rel}}' =   - \mathbf{N}  \cdot   \vct v_{\mathrm{rel}} \,,  \qquad \hat{\mathbf{z}} \cdot     \left( \mathbf{N} \times  \vct v_{\mathrm{rel}}' \right)   = \hat{\mathbf{z}} \cdot     \left( \mathbf{N} \times  \vct v_{\mathrm{rel}} \right)  \, . 
\end{align}
The total relative velocity $\vct v_{\mathrm{rel}}$ is given by 
\begin{align}
     \vct v_{\mathrm{rel}}  = \mathbf{v}_1 - \mathbf{v}_2  +  \hat{\mathbf{z}} \times \mathbf{r}_1 \omega_1  -  \hat{\mathbf{z}} \times \mathbf{r}_2 \omega_2 \, , 
\end{align}
\end{subequations}
where $\mathbf r_{1,2}$ is the distance vector of the collision point from the respective particles. From \eqref{eq:Nvector}, the outgoing linear velocities can be computed to be
\begin{align}
\mathbf{v}_{1,2}'  =    \mathbf{v}_{1,2}  \mp   \mathbf{N} \frac{   \mathbf{N}  \cdot \vct v_{\mathrm{rel}}}{1  +  m \frac{ \ell^2_1   +\ell_2^2 }{2 I }    }   ~~ , 
\end{align}
where 
\begin{align}
 \ell_{1,2} =  \mathbf{N} \cdot  \hat{\mathbf{z}} \cross \mathbf r_{1,2}  
\end{align}
In order to compute the normal vector $\mathbf{N}$ we select the ratchet whose collision point is not at one of its pointy ends, as there $\mathbf{N}$ becomes ambiguous. It follows from \eqref{eq:boundarycoord} that 
\begin{align}
   \mathbf{N}_{\text{unnorm}} =   \pm    \left[  \hat{\mathbf{r}}_{1,2} -   \frac{\partial_{\phi } r_{1,2} ( \phi ) }{ r_{1,2} ( \phi ) } \hat{\bm{\phi}}  \right]   ~~  . 
\end{align}
After $\mathbf{N}_{\text{unnorm}}$ has been extracted, it is converted to Cartesian components and normalized to obtain $\mathbf{N}$. Through $\mathbf{v}_{1,2}'$ we obtain the scattering angle $\chi$. 
Then, for every sampled impact parameter $b_j$, we perform
$N_{\mathrm{trials}}$ independent trials. We divide the full range
of scattering angles into bins $B_n$ of width $\Delta\chi$, and
denote the center of each bin by $\chi_n$. Let $N_n(b_j)$ be the
number of trials at impact parameter $b_j$ for which a collision
occurs and the resulting scattering angle lies in $B_n$. The
differential cross section is estimated as
\begin{equation}
\frac{\partial \sigma_{\mathrm{ratchet}}}{\partial \chi}(\chi_n)
\approx
\sum_j
\frac{\Delta b}{\Delta \chi}
\frac{N_n(b_j)}{N_{\mathrm{trials}}}.
\end{equation}
The sampled impact parameters cover the range
\begin{equation}
-2R_{\mathrm{outer}}
\leq b
\leq
2R_{\mathrm{outer}},
\end{equation}
which contains all impact parameters for which a collision between
the two ratchets is geometrically possible. The differential cross section based on hard disks with chiral effective radii is given by 
\[
\begin{aligned}
\frac{\partial \sigma_{\varepsilon} (\chi)}{\partial  \chi }  &= r (1 + \varepsilon) \sin(\chi/2 )    &&   0  <  \chi <   \pi  ,  \\
\frac{\partial \sigma_{\varepsilon} (\chi)}{\partial  \chi }  &=   r (1   -  \varepsilon) \sin(\chi/2 )  , &&   \pi   <  \chi <   2 \pi   .  
\end{aligned}
\]
We fit $\varepsilon$ and $r$ with the least squares method in order to best capture the differential cross section of the ratchet collision experiment with the chiral effective radius, which leads to the histogram shown in Fig.~\ref{fig:placeholder2}.

\section*{$H$-theorem}
\label{app:h_theorem}
To show that $H$-theorem holds, it is helpful to write the collision integral as
    \begin{align}
\begin{split}
 &      C[f  (\mathbf{v}_1 )] = \iiint   d \mathbf{v}_2 d \mathbf{v}'_2 d \mathbf{v}'_1  \left(    W(  \vct v_1 ,\vct v_2  | \vct v'_1,\vct v'_2 ) f (\mathbf{v}'_1 ) f (\mathbf{v}'_2)   -    W(\vct v_1',\vct v_2' | \vct v_1,\vct v_2) f  (\mathbf{v}_1) f (\mathbf{v}_2)  \right)     ,  \end{split}
\end{align}
with the transition probability $ W(\vct v'_1, \vct v'_2  | \vct v_1, \vct v_2)$ given by
\begin{align}  \label{eq:Wfunctoin}
\begin{split}
       &   W(\vct v'_1, \vct v'_2  | \vct v_1, \vct v_2) =  |\vct g  |   \sum_{ \pm}     \frac{d (1 \pm \varepsilon )}{2}    \int_{ \frac{\pi}{2}   \mp  \frac{\pi}{2} }^{\frac{3 \pi}{2}   \mp \frac{\pi}{2} } d\chi \sin\left(\frac{\chi}{2}\right)  \delta^{(2)}\left[\vct v_1'-\vct v_1'(\vct v_1, \vct v_2,\chi)\right]\delta^{(2)}\left[\vct v_2'-\vct v_2'(\vct v_1, \vct v_2,\chi)\right],
\end{split}    
\end{align}
where the primed velocities are equal to
\begin{align}
        \vct v_{1,2} '(\vct v_1, \vct v_2,\chi) &= \frac{\vct v_1+\vct v_2}{2} \pm \mathbf{R}( \chi ) \cdot \frac{\vct v_1-\vct v_2}{2}, 
\end{align}
as we infer from Eq.~(\ref{eq:rottttt}).
$H$-theorem is typically derived by utilizing microscopic reversibility \cite{degroot1984nonequilibrium,Boltzmann1964}. However, the chirality of the collisions in Fig.~\ref{fig:placeholder} means that $W$ is invariant under neither parity nor time-reversal:
        \begin{align} \label{eqapp:coll_inequal}
        W( \vct v_1',\vct v_2' | \vct v_1,\vct v_2)  \neq  W( \vct v_1,\vct v_2 | \vct v'_1,\vct v'_2) \, . 
    \end{align}
Despite this, we can still prove the $H$-theorem 
\begin{align}  \label{eq:Htheorem}
    \frac{\partial H}{ \partial t} \leq 0 \, , 
\end{align}
where the $H$-functional is given by
\begin{align}  
    H = \iint d \mathbf{r}  d \mathbf{v}_1   f (\mathbf{v}_1 )  \log(f (\mathbf{v}_1 ) ) ~~ . 
\end{align}
To prove that \eqref{eq:Htheorem} holds, let us write
\begin{align}
        \frac{\partial H}{ \partial t} = \int d  \mathbf{r} \, Q ~~ . 
\end{align}
It follows from \eqref{eq:fequation} that
\begin{align}  \label{eq:integrallll}
\begin{split}
     &    Q  
=  \idotsint d\mathbf{v}_1' \, d\mathbf{v}_2' \, d\mathbf{v}_1 \, d\mathbf{v}_2  
 \log (  f (\mathbf{v}_1 )  )    
\left(    W(  \vct v_1 ,\vct v_2  | \vct v'_1,\vct v'_2 )   f (\mathbf{v}'_1)  f (\mathbf{v}'_2)   -    W(\vct v_1',\vct v_2' | \vct v_1,\vct v_2) f  (\mathbf{v}_1 ) f (\mathbf{v}_2)  \right)     .
\end{split}
\end{align}
Since the velocities in \eqref{eq:integrallll} are dummy variables, it holds that
\begin{align}  \label{eq:intell}
\begin{split}
     &  Q 
 =  \frac{1}{2}  
\idotsint d\mathbf{v}_1' \, d\mathbf{v}_2' \, d\mathbf{v}_1 \, d\mathbf{v}_2  \,
\log\!\left( \frac{ f(\mathbf{v}'_1)\, f(\mathbf{v}'_2) }{f(\mathbf{v}_1)\, f(\mathbf{v}_2)  } \right)    W (\mathbf{v}_1 , \mathbf{v}_2 | \mathbf{v}'_1 , \mathbf{v}'_2  )  f(\mathbf{v}_1')\, f(\mathbf{v}_2')   
.
\end{split}
\end{align}
Using \eqref{eq:Wfunctoin}, we can calculate
\begin{align}
    \begin{split}
         &     \int d^2v'_1 d^2v'_2 W( \mathbf{v}'_1 , \mathbf{v}'_2  | \mathbf{v}_1 , \mathbf{v}_2  )  =2d|\vct g |  ~~ , ~~  \int d^2v_1 d^2v_2 W( \mathbf{v}'_1 , \mathbf{v}'_2  | \mathbf{v}_1 , \mathbf{v}_2  )  =2d|\vct g' |.  \end{split}
\end{align}
From \eqref{eq:conservation} we learn that
\begin{align}
    |\vct g | = |\vct g' | \,.
\end{align}
It thus follows that
\begin{align}  \label{eq:integrallll12cccccc}
\begin{split}
 & \idotsint d\mathbf{v}_1' \, d\mathbf{v}_2' \, d\mathbf{v}_1 \, d\mathbf{v}_2 
  W (  \mathbf{v}'_1 , \mathbf{v}'_2 |  \mathbf{v}_1 , \mathbf{v}_2)  f(\mathbf{v}_1')\, f(\mathbf{v}_2')  =   \idotsint  d\mathbf{v}_1' \, d\mathbf{v}_2' \, d\mathbf{v}_1 \, d\mathbf{v}_2  W ( \mathbf{v}'_1 , \mathbf{v}'_2  | \mathbf{v}_1 , \mathbf{v}_2  ) f(\mathbf{v}_1)\, f(\mathbf{v}_2) 
\, , 
\end{split}
\end{align}
which can be rewritten as
\begin{align}  \label{eq:integral233}
\begin{split}
   &  \idotsint  d\mathbf{v}_1' \, d\mathbf{v}_2' \, d\mathbf{v}_1 \, d\mathbf{v}_2   W (   \mathbf{v}'_1 , \mathbf{v}'_2  |    \mathbf{v}_1 , \mathbf{v}_2)  f(\mathbf{v}_1)\, f(\mathbf{v}_2)  \left(  \frac{ f(\mathbf{v}'_1)\, f(\mathbf{v}'_2) }{  f(\mathbf{v}_1)\, f(\mathbf{v}_2)  }    -1  \right)   =0 
~~ . 
\end{split}
\end{align}
Using \eqref{eq:integral233}, \eqref{eq:intell} can be written as
\begin{align}  \label{eq:intell123213}
\begin{split}
     &  Q 
 =  \frac{1}{2}
\idotsint d\mathbf{v}_1' \, d\mathbf{v}_2' \, d\mathbf{v}_1 \, d\mathbf{v}_2  \,
\left[ \log\!\left( \frac{ f(\mathbf{v}'_1)\, f(\mathbf{v}'_2) }{f(\mathbf{v}_1)\, f(\mathbf{v}_2)  } \right)   -  \frac{ f(\mathbf{v}'_1)\, f(\mathbf{v}'_2) }{  f(\mathbf{v}_1)\, f(\mathbf{v}_2)  }    + 1   \right]  W (  \mathbf{v}'_1 , \mathbf{v}'_2  |  \mathbf{v}_1 , \mathbf{v}_2 )  f(\mathbf{v}_1)\, f(\mathbf{v}_2) ~~  .  \end{split}
\end{align}
Since it holds that \cite{Cercignani1981}
\begin{align}
    \log(x) -x +1 \leq 0 ~~ , ~~ x \geq 0   
\end{align}
it follows that
\begin{align}
    Q \leq 0  ~~ , 
\end{align}
so that \eqref{eq:Htheorem} is proven.

\section*{Sonine polynomial expansion}
\label{app:invert}
In this section we explain how to solve the linearized Boltzmann equation \eqref{eq:Phi} for $\Phi$ and thereby extract transport viscosity and thermal conductivity. Strictly speaking, the solution to \eqref{eq:Phi} is not unique since $C^{(1)}$ has a kernel spanned by the functions $1$, $c_i$ and $\vct c^2$ (because collisions conserve particle number, linear momentum and kinetic energy), but as is customary, we make $\Phi$ unique by defining the inner product in the space of complex functions
\begin{equation}
    (\Phi,\Psi) = \int d \vct v_1 f_0(\vct v_1)\Phi(\vct v_1)\Psi(\vct v_1)
    \label{eq:inner_product12}
\end{equation}
and demanding that 
\begin{align}
    (\Phi,1)=(\Phi,c_i)=(\Phi,\vct c^2)=0 ~~ .  \label{labeldemands123} 
\end{align}
The left-hand side of the Boltzmann equation~\eqref{eq:Phi} reads explicitly \cite{Dorfman}: 
\begin{equation}
    \mathcal{L}[f_0] = f_0(\vct v, \vct x, t)\left[\frac{m}{2\kb T}(c_ic_j-\frac{1}{2}\delta_{ij}c^2)\left(\partial_iu_j+\partial_ju_i\right)+\left(\frac{mc^2}{2\kb T}-2\right)c_i   \partial_i  \log T  \right],
    \label{eq:boltz_lhs}
\end{equation}
where $  \vct c \equiv \vct v - \vct u(\vct x,t)$ is called the peculiar velocity and we used the Einstein summation convention. 
From \eqref{eq:boltz_lhs} it follows that the solution has the form
\begin{align}
\begin{split}
     \Phi     & =   \sum_{(\even , \odd) }  \left( - \frac{1}{n} \sqrt{ \frac{ 2 \kb T }{m}} \mathbf{ A }^{(\even , \odd )}  \cdot \nabla  \log  T  -  \frac{2}{n}  \mathbf{B}^{(\even , \odd )}  :\nabla  \mathbf{u}  \right)   ~~  ,  
     \end{split}
     \label{eqapp:phi_ansatz}
\end{align}
where $\vct A^{(\even , \odd )}$ and $\vct B^{(\even , \odd )}$ are some functions of $\vct c$ to be determined. The tensors labeled $(e)$ are parity even, while the ones labeled $(o)$ are parity odd.
Defining $   C^{(1)} [ \Psi ]  = - n^2 I^{(1)} [ \Psi ] $, where for a general field $\Psi$ we have
\begin{align}  \label{eq:Cexpnded12222}
    I^{(1)} \left[ \Psi \right]  =   \frac{d }{ 2 n^2}   \sum_{ \pm}      (1 \pm \varepsilon )  \int_{ \frac{\pi}{2}   \mp  \frac{\pi}{2} }^{\frac{3 \pi}{2}   \mp \frac{\pi}{2} } d \chi \sin(\chi /2  )  \int  d \mathbf{v}_2  |\mathbf{g} |   f_0(\vct v)  f_0(\vct v_2)  \left[\Psi(\vct v)+\Psi(\vct v_2)-\Psi(\vct v_1')-\Psi(\vct v_2')\right]  ~~  ,   \end{align} 
we can write \eqref{eq:Phi} as 
\begin{subequations}  \label{eq:masterequation}
    \begin{align}   \label{eq:masterequation1}
 &   ( \mathbf{X} \mathbf{X} -    \mathbf{1} X^2  /2   )  f_0(\vct v)  =        n         \sum_{( \even  ,\odd )}   I^{(1)} [ \mathbf{B}^{( \even  ,\odd )}   ]\,, 
\end{align}
\begin{align}  
 &   \left(   X^2     -2 \right)   \mathbf{X}  f_0(\vct v)  =    n    \sum_{( \even  ,\odd )} I^{(1)} [ \mathbf{A}^{( \even  ,\odd )}     ]\,, 
 \label{eq:masterequation2}
 \end{align}
\end{subequations}
where we introduced the dimensionless velocity
\begin{align} \label{eq:Xdefinition}
    \mathbf{X} =  \sqrt{ \frac{m}{ 2 \kb T }}  \mathbf{c}\, . 
\end{align}
To solve \eqref{eq:masterequation}, we expand $\mathbf A^{( \even  ,\odd )}$ and $\mathbf B^{( \even  ,\odd )}$ in the basis of the Sonine polynomials $S_n^{(m)}$, also known as the generalized Laguerre polynomials. The Sonine polynomials satisfy the following identity involving the inner product in Eq.~(\ref{eq:inner_product12}):
\begin{equation}
    (  S^{(m)}_{n}, X^{ 2 m}S^{(m)}_{n'}) = \delta_{nn'}\frac{(n+m)!}{n!} .
    \label{eqapp:orthonormal}
\end{equation}
Our expansion reads
\begin{subequations}
\begin{align}
\mathbf{B}^{(\even , \odd )}  & = \sum^{N}_{n = 0}  \mathbb b^{(\even , \odd )}_n  \mathbf{b}^{(\even , \odd )}_{n}\,, \qquad \mathbf{b}^{(\even , \odd )}_n  =    \mathbf{X} \mathbf{X} : \mathbf{T}^{(\even , \odd )}  S_n^{(2)} (X^2 )   \end{align}
and 
\begin{align} 
\label{eq:Asonine}
\mathbf{A}^{(\even , \odd )}  & = \sum^{N+1}_{n = 1}  \mathbb a^{(\even , \odd )}_n  \mathbf{a}^{(\even , \odd )}_{n} ~~ , ~~ \mathbf{a}^{(\even , \odd )}_n  =   \mathbf{X} \cdot  \mathbf{P}^{(\even , \odd)}  S_n^{(1)} (X^2 )   ~~ , 
\end{align}
\end{subequations}
where $\mathbb b^{(\even , \odd )}_n$ and $\mathbb a^{(\even , \odd )}_n$ are constants and we introduced the tensors 
\begin{align}
    \begin{split}
    P_{ij}^{(e)}   =  \delta_{ij }\,, ~~  P_{ij}^{(o)}  =  \epsilon_{ij } \, ,  ~~  
      T_{ijkl}^{(e)}    = \frac{1}{2} \left(  \delta_{ik } \delta_{jl} + \delta_{jk } \delta_{il} - \delta_{i j  } \delta_{kl } \right) \,, ~~   
      T_{ijkl}^{(o)}  = \frac{1}{2}  ( \delta_{i k }  \epsilon_{jl } + \delta_{j l } \epsilon_{ik }      )  \, . 
    \end{split}
\end{align}
The tensors $\vct P^{(e,o)}$ are the only two-component tensors consistent with isotropy in two dimensions, and likewise $T^{(e,o)}$ are the only isotropic four-component tensors symmetric and traceless in the last two indices, related to the fact that only the traceless part is nonzero on the left hand side of \eqref{eq:masterequation1}. We excluded $S_0^{(1)} (X^2 ) $ in the sum of \eqref{eq:Asonine} to satisfy \eqref{labeldemands123}.
Using Eq.~(\ref{eqapp:orthonormal}), we derive from \eqref{eq:masterequation1} the relations 
\begin{subequations}  
\label{eq:coefficients12312}
    \begin{align}
     \sum^N_{m =0 }   \int d  \mathbf{v}     \mathbf{b}^{(\even)}_n  :  I^{(1) } \left[ \mathbb b^{(\even)}_m  \mathbf{b}^{(\even)}_m  + \mathbb b^{(\odd)}_m  \mathbf{b}^{(o)}_m  \right] \equiv   \mathbf{b}_N \cdot  \mathbb b^{(\even)}  +    \mathbf{b}_N^{\prime  } \cdot \mathbb b^{(\odd)}   &  = \frac{1}{2} \delta_{ n 0 } \, ,   \\ 
       \sum^{N}_{m =0 }   \int d  \mathbf{v}    \mathbf{b}^{(o)}_n  :  I^{(1) } \left[ \mathbb b^{(\even)}_m  \mathbf{b}^{(\even)}_m  + \mathbb b^{(\odd)}_m  \mathbf{b}^{(o)}_m  \right] \equiv   \mathbf{b}_N^* \cdot  \mathbb b^{(\even)}  +    \mathbf{b}_N^{\prime *   } \cdot \mathbb b^{(\odd)}   &   = 0 \, , \end{align}
  \end{subequations}
whereas we find from \eqref{eq:masterequation2} the relations
\begin{subequations}  
\label{eq:coefficients123121}
    \begin{align}
      \sum^{N+1}_{m =1 }   \int d  \mathbf{v}    \mathbf{a}^{(\even )}_n  \cdot   I^{(1) } \left[ \mathbb a^{(\even)}_m  \mathbf{a}^{(\even)}_m  + \mathbb a^{(\odd)}_m  \mathbf{a}^{(o)}_m  \right] \equiv \mathbf{a}_N \cdot  \mathbb a^{(\even)}  +    \mathbf{a}_N^{\prime  } \cdot \mathbb a^{(\odd)}     &  = 2 \delta_{n0} ,   \\ 
     \sum^{N+1}_m   \int d  \mathbf{v}    \mathbf{a}^{(o)}_n  \cdot   I^{(1) } \left[ \mathbb a^{(\even)}_m  \mathbf{a}^{(\even)}_m  + \mathbb a^{(\odd)}_{m =1 }   \mathbf{a}^{(o)}_m  \right] \equiv \mathbf{a}_N^* \cdot  \mathbb a^{(\even)}  +    \mathbf{a}_N^{\prime *   } \cdot \mathbb a^{(\odd)}     &  = 0 \, . \end{align}
  \end{subequations}
The out-of-equilibrium stress tensor $\sigma_{ij}$ and the heat current $q_i$ are
\begin{align}
\sigma_{ij}  = ( m c_i c_j,\Phi)~~ , ~~      q_i  = \left(\frac{m c^2}{2}c_i,\Phi\right) \label{eq:stress_heat1} ~~ . 
\end{align} 
The transport coefficients enter into these currents as
\begin{subequations}
\begin{align}
    \sigma_{ij} & = - 2 \sum_{(e,o)}\eta_{e,o}T^{(e,o)}_{ijkl} \partial_k u_l,  \\
        q_i & = -\sum_{(e,o)}\kappa_{e,o}P^{(e,o)}_{ij}\partial_j  \log  T,
\end{align} \label{eq:trans_coeffs_def}
\end{subequations}
Therefore, we find the following formulae for the transport coefficients:
\begin{subequations}
    \begin{align}
    \label{eq:etarelation}
    \eta^{(N)}_{\even ,  \odd  }  &  = \mathbb b^{(\even ,  \odd  )}_0     \kb T      \\ 
    \kappa^{(N)}_{\even ,  \odd  } &  = \mathbb  a^{(\even ,  \odd  )}_1  \frac{    ( \kb T )^2 }{m  }         
\end{align}
\end{subequations}
From \eqref{eq:coefficients12312} it follows that
\begin{subequations} \label{dexpression12311}
    \begin{align}  
   \eta^{(N)}_{\even    }   & = \frac{1}{2} \kb T  \left(  \mathbf{b}_N  +  \mathbf{b}_N^{\prime } (\mathbf{b}_N  )^{-1}   \mathbf{b}_N^{\prime  } \right)^{-1}_{00}  ~~ ,   \\ 
        \eta^{(N)}_{  \odd  } &  =  \frac{1}{2} \kb T  \left( \mathbf{b}_N^{\prime }  +  \mathbf{b}_N (\mathbf{b}_N^{\prime } )^{-1}   \mathbf{b}_N \right)^{-1}_{00}  ~~  , 
\end{align}
\end{subequations}
  where we used that
\begin{subequations}
    \begin{align}
    \mathbf{b}_N   &  = \mathbf{b}_N^{* \prime } ~~ , ~~ 
    \mathbf{b}_N^{\prime }  =   - \mathbf{b}_N^{*} ~~ . 
\end{align}
\end{subequations}
Similarly, we find from \eqref{eq:coefficients123121} that
\begin{subequations} \label{dexpression12511}
    \begin{align}  
   \kappa^{(N)}_{\even    }   & =    \frac{ 2    ( \kb T )^2 }{m  } \left(  \mathbf{a}_N  +  \mathbf{a}_N^{\prime } (\mathbf{a}_N  )^{-1}   \mathbf{a}_N^{\prime  } \right)^{-1}_{11}  ~~ ,   \\ 
        \kappa^{(N)}_{  \odd  } &  =    \frac{  2    ( \kb T)^2  }{m  }  \left( \mathbf{a}_N^{\prime }  +  \mathbf{a}_N (\mathbf{a}_N^{\prime } )^{-1}   \mathbf{a}_N \right)^{-1}_{11}  ~~  , 
\end{align}
\end{subequations}
  where we used that
\begin{subequations}
    \begin{align}
    \mathbf{a}_N   &  = \mathbf{a}_N^{* \prime } ~~ , ~~ 
    \mathbf{a}_N^{\prime }  =   - \mathbf{a}_N^{*} ~~ . 
\end{align}
\end{subequations}
Lastly, we have
\begin{subequations} \label{dexpression123}
    \begin{align}  
   N^{(N)}_{\even    } (\varepsilon)  & =  \left( b_{0}  + \frac{(b_{0}^{\prime })^2}{b_{0}  } \right)   \left(  \mathbf{b}_N  +  \mathbf{b}_N^{\prime } (\mathbf{b}_N  )^{-1}   \mathbf{b}_N^{\prime  } \right)^{-1}_{00}  ~~ ,   \\ 
        N^{(N)}_{  \odd  }  (\varepsilon) &  =  \left( b'_{0}  + \frac{b_{0}^2}{b^{\prime }_{0}  } \right)  \left( \mathbf{b}_N^{\prime }  +  \mathbf{b}_N (\mathbf{b}_N^{\prime } )^{-1}   \mathbf{b}_N \right)^{-1}_{00}  ~~  , 
\end{align}
\end{subequations}
 and 
\begin{subequations} \label{dexpression125}
    \begin{align}  
   K^{(N)}_{\even }  (\varepsilon)   & =  \left( a_{1}  + \frac{(a_{1}^{\prime })^2}{a_{1}  } \right)  \left(  \mathbf{a}_N  +  \mathbf{a}_N^{\prime } (\mathbf{a}_N  )^{-1}   \mathbf{a}_N^{\prime  } \right)^{-1}_{11}  ~~ ,   \\ 
        K^{(N)}_{\odd }  (\varepsilon)  &  =   \left( a'_{1}  + \frac{a_{1}^2}{a^{\prime }_{1}  } \right) \left( \mathbf{a}_N^{\prime }  +  \mathbf{a}_N (\mathbf{a}_N^{\prime } )^{-1}   \mathbf{a}_N \right)^{-1}_{11}  ~~  , 
\end{align}
\end{subequations}
which gives \eqref{eq:finalresult2}. 

To better understand how the elements of the matrices in \eqref{dexpression12311} and \eqref{dexpression12511} are computed, let us explicitly work out $b_0 $ and $b'_0$, which allow for the determination of $ \eta^{(0)}_{\even    } $ and $ \eta^{(0)}_{\odd    } $. Using \eqref{eq:Cexpnded12222}, we find
\begin{subequations} \label{eq:Cexpnded1133333}
    \begin{align}  
    \begin{split}
             & b_0   =  
    \frac{d }{  \pi^2  } \sqrt{\frac{\kb T }{2 m }}   \sum_{ \pm}      (1 \pm \varepsilon ) \cdot   \\  &   \cdot \int_{ \frac{\pi}{2}   \mp  \frac{\pi}{2} }^{\frac{3 \pi}{2}   \mp \frac{\pi}{2} } d \chi \sin(\chi /2  )  \iint  d \mathbf{X} d \mathbf{X}_2  |\mathbf{X}  - \mathbf{X}_2  | \exp(- X^2 - X_2^2 )   \left[\mathbf{X} \mathbf{X}    +\mathbf{X}_2 \mathbf{X}_2 -\mathbf{X}'_1 \mathbf{X}'_1 -\mathbf{X}'_2 \mathbf{X}'_2\right]  :  \mathbf{T}^{(\even)} :  \mathbf{X} \mathbf{X}   ~~  ,   
        \end{split}   \\ 
          \begin{split}
             & b'_0   =  
    \frac{d }{  \pi^2  } \sqrt{\frac{\kb T }{2 m }}   \sum_{ \pm}      (1 \pm \varepsilon ) \cdot   \\  &   \cdot \int_{ \frac{\pi}{2}   \mp  \frac{\pi}{2} }^{\frac{3 \pi}{2}   \mp \frac{\pi}{2} } d \chi \sin(\chi /2  )  \iint  d \mathbf{X} d \mathbf{X}_2  |\mathbf{X}  - \mathbf{X}_2  | \exp(- X^2 - X_2^2 ) \left[\mathbf{X} \mathbf{X}    +\mathbf{X}_2 \mathbf{X}_2 -\mathbf{X}'_1 \mathbf{X}'_1 -\mathbf{X}'_2 \mathbf{X}'_2\right]  : \mathbf{T}^{(\odd )} : \mathbf{X} \mathbf{X}  ~~ .   
        \end{split}
\end{align}
\end{subequations}
Momentum conservation during the collision allows us to parametrize the variables in \eqref{eq:Cexpnded1133333} as 
    \begin{align}
    \mathbf{X}  &= \mathbf{X}_G + \frac{1}{2} \mathbf{X}_g      ~~  , ~~  \mathbf{X}_2    = \mathbf{X}_G - \frac{1}{2} \mathbf{X}_g     ~~  , ~~  \mathbf{X}_1'     =\mathbf{X}_G +  \frac{1}{2} \mathbf{X}_g'   
~~  , ~~
    \mathbf{X}_2'    =  \mathbf{X}_G - \frac{1}{2} \mathbf{X}_g'   ~~ ,  
\end{align}
so that \eqref{eq:Cexpnded1133333} can be rewritten as
\begin{subequations} \label{eq:Cexpnded11333334}
    \begin{align}  
    \begin{split}
             & b_0   =  
    \frac{d }{  \pi^2  } \sqrt{\frac{\kb T }{2 m }}   \sum_{ \pm}      (1 \pm \varepsilon ) \cdot \\  &  
    \cdot  \int_{ \frac{\pi}{2}   \mp  \frac{\pi}{2} }^{\frac{3 \pi}{2}   \mp \frac{\pi}{2} } d \chi \sin(\chi /2  )  \iint  d \mathbf{X}_G d \mathbf{X}_g  |\mathbf{X}_g  | \exp(- \frac{1}{2} X_g^2 - 2 X_G^2 ) \left[\mathbf{X} \mathbf{X}    +\mathbf{X}_2 \mathbf{X}_2 -\mathbf{X}'_1 \mathbf{X}'_1 -\mathbf{X}'_2 \mathbf{X}'_2\right]  : \mathbf{T}^{(\even)} :   \mathbf{X} \mathbf{X}  ~~  ,  
        \end{split}  \\
          \begin{split}
             & b'_0   =  
    \frac{d }{  \pi^2  } \sqrt{\frac{\kb T }{2 m }}   \sum_{ \pm}      (1 \pm \varepsilon ) \cdot \\  &  
    \cdot  \int_{ \frac{\pi}{2}   \mp  \frac{\pi}{2} }^{\frac{3 \pi}{2}   \mp \frac{\pi}{2} } d \chi \sin(\chi /2  )  \iint  d \mathbf{X}_G d \mathbf{X}_g  |\mathbf{X}_g  | \exp(- \frac{1}{2} X_g^2 - 2 X_G^2 )  \left[\mathbf{X} \mathbf{X}    +\mathbf{X}_2 \mathbf{X}_2 -\mathbf{X}'_1 \mathbf{X}'_1 -\mathbf{X}'_2 \mathbf{X}'_2\right] : \mathbf{T}^{(\odd )} :  \mathbf{X} \mathbf{X}  ~~   , 
        \end{split}
\end{align}
\end{subequations}
From conservation of energy given by \eqref{eq:energetic}, it follows that $    \mathbf{X_g}' = | \mathbf{X_g}| \left( \cos(\chi) , \sin(\chi ) \right) $ when $ \mathbf{X_g} = | \mathbf{X_g} | \left( 1  ,0 \right) $, which for this integral we can assume without loss of generality. With this information, we can straightforwardly work out the integral of \eqref{eq:Cexpnded11333334}, which yields
\begin{align}
    b_0 = d    \sqrt{ \frac{   \pi   \kb T } {m}}     ~~ , ~~ b_0'  =-   \frac{d  \epsilon }{4}        \sqrt{ \frac{  \pi   \kb T } {m}}   ~~ , 
\end{align}
from which the first line of \eqref{eq:finalresult} follows when plugged into \eqref{dexpression12311} for $N=0 $. A similar calculation allows us to arrive at
\begin{align}
    a_1 = d    \sqrt{ \frac{   \pi   \kb T } {m}}     ~~ , ~~ a_1'  =-   \frac{d  \epsilon }{4}        \sqrt{ \frac{  \pi   \kb T } {m}}   ~~ , 
\end{align}
which plugged into~\eqref{dexpression12511} gives the second line of \eqref{eq:finalresult}.

\section*{Nonequilibrium molecular dynamics}
\label{app:NEMD}
In this section we describe the nonequilibrium molecular dynamics computations that we perform to verify \eqref{eq:finalresult}. In the simulation, we consider a shear periodic box of hard disks which collide with the collision rule of Fig.~\ref{fig:placeholder}. We turn on shear using the SLLOD algorithm \cite{Evans2007-dg,PhysRevA.30.1528,10.1063/1.2192775,han2021fluctuating}, which means that particles evolve according to 
\begin{align}  \label{eq:evolutionequations}
\begin{split}
        \dot{x}_i    & =   u_{ i} +  \gamma  y_i    , \qquad     
        \dot{y}_i    = v_{ i}   \, ,    \\ 
    \dot{u}_{ i}   & =  \frac{1}{m} \sum_{j \neq i} f^x_{ij}    - \gamma  v_{ i }   + \alpha u_i    \, , \quad   
\dot{v}_{ i}    =  \frac{1}{m} \sum_{j \neq i} f^y_{ij}  + \alpha v_i     \,   , 
\end{split}
\end{align}
where $\gamma$ is the shear rate and $\mathbf{f}_{i j}$ is the force on particle $i$ due to collisions with particle $j$.
The coefficient $\alpha$ is related to a thermostat mechanism introduced to prevent the system from heating up, which will be discussed shortly. We consider shear $\gamma$ that acts with a constant magnitude during a single run, but its sign changes periodically, with the period long enough that the gas has enough time to thermalize. To extract viscosity, we compute
\begin{align}
    \eta = \left\langle  \frac{\sigma_{xy}}{\gamma }    \right\rangle \,, \quad     \eta_\odd  = \frac{1}{2 } \left\langle  \frac{\sigma_{xx } - \sigma_{yy }}{\gamma }    \right\rangle \, , 
\end{align}
where $\langle  ... \rangle $ corresponds to an average over the run and $\bm{\sigma} $ is the Irving-Kirkwood stress given by
\begin{equation}  \label{eq:irvingkirkwood}
\bm{\sigma}  = -\frac{1}{A} 
\left[
\sum_i m \mathbf{v}_i \mathbf{v}_i 
+ \frac{1}{2} \sum_{i \neq j} \mathbf{f}_{ij} \mathbf{r}_{ij}
\right] \,  , 
\end{equation}
where $A$ is the box area. Because of the simple hard disk nature of the collisions, the second term on the right hand side of \eqref{eq:irvingkirkwood} reduces to
\begin{align}
     \frac{1}{2} \sum_{i \neq j} \mathbf{f}_{ij} \mathbf{r}_{ij}  =  m  \sum_{n_c} \frac{  \Delta \mathbf{v}_{n_c} }{\Delta t }  \Delta \mathbf{x}_{n_c} \,, 
\end{align}
where $\Delta \mathbf{v}_{n_c} $ is the change of velocity of the colliding particles caused by the $n_c$th collision of a time step of the numerical simulation and $\Delta t$ is the width of this time step. Note that because $ \Delta \mathbf{v}_{n_c} ||  \Delta    \mathbf{x}_{n_c}$, it holds that $\bm{\sigma}= \bm{\sigma}^T $, confirming that the chiral collisions considered in this work conserve angular momentum. 

\begin{figure}
    \centering
    \includegraphics[width=0.5\linewidth]{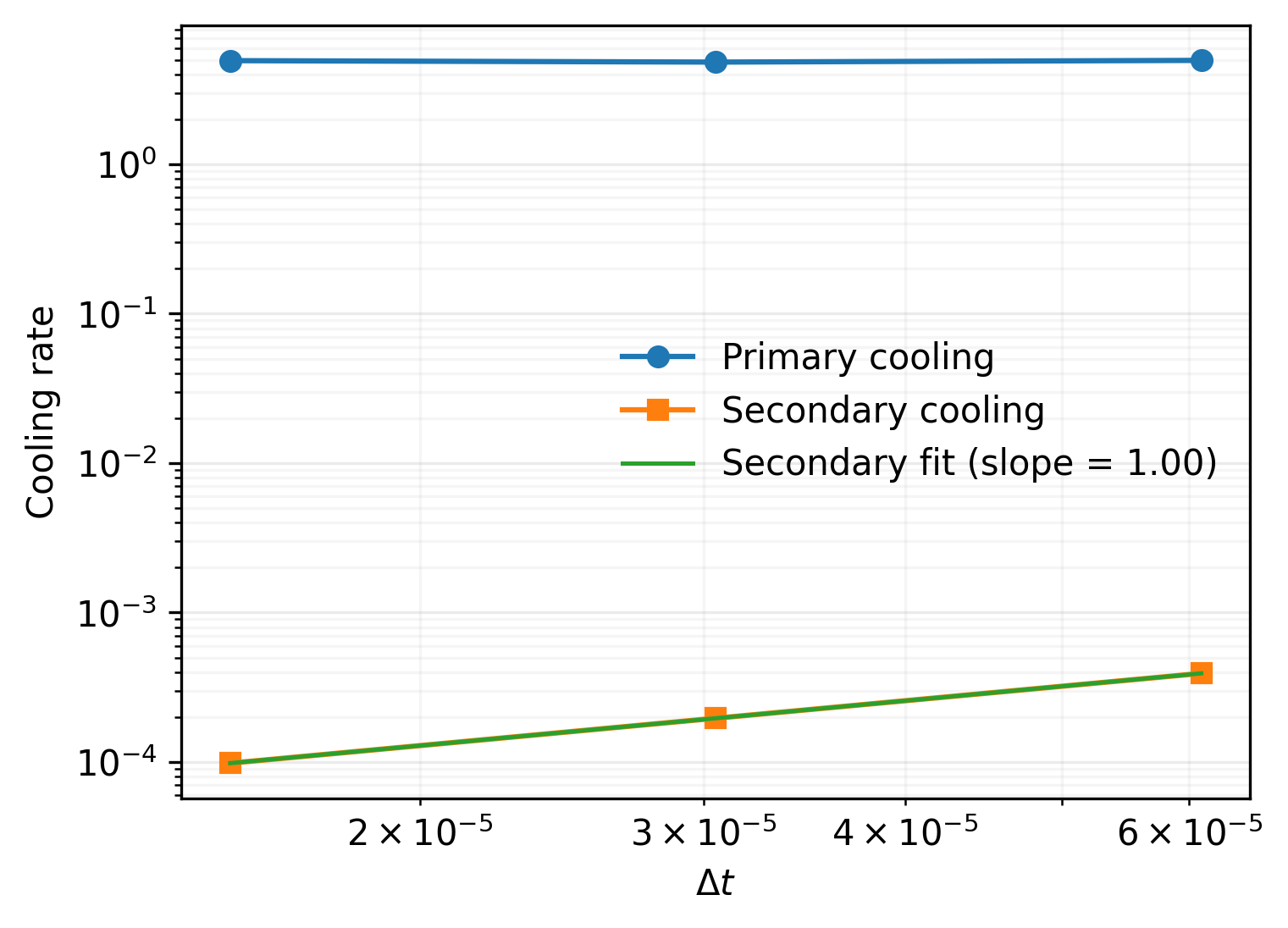}
    \caption{Cooling power associated with the primary Gaussian
    isokinetic thermostat and the secondary numerical correction as a
    function of the timestep $\Delta t$.  The primary cooling remains
    approximately constant under temporal refinement, while the
    secondary cooling scales linearly with $\Delta t$, demonstrating
    that it is a finite-timestep numerical effect.}
    \label{fig:thermostat_convergence}
\end{figure}

The shear induced by $\gamma$ heats up the system, which is problematic because the transport coefficients we wish to compute depend on temperature. To remedy this, we use a thermostat that controls the temperature. In particular, we choose $\alpha$ in \eqref{eq:evolutionequations} in such a way that total momentum is conserved and the added kinetic energy due to shearing is canceled out, i.e.  
\begin{align}
\begin{split}
   &    \frac{1}{2}  m    \frac{\partial }{\partial  t }  \sum_{i} (u_i^2 + v_i^2 )      =   -  m  \sum_{i}   \left( \gamma   v_i u_i     - \alpha ( u_i^2 + v_i^2 )   \right)  =0\,,
     \end{split}
\end{align}
which requires that \cite{Evans2007-dg}
\begin{align}
    \alpha   =  \frac{  \gamma  \sum_{i}  v_i u_i  }{\sum_{i} ( v_i^2 +  u^2_i ) } ~~ .  
\end{align}
After cooling the system down with this thermostat, there is still a tiny amount of heating which adds up after a large number of time steps. To eliminate this residual heating effect, we additionally cool down the system by subtracting the mean velocity and rescaling the velocities. \textcolor{black}{To verify that this secondary cooling is a numerical rather than a physical effect, we performed a single back-and-forth cycle at the temporal resolution used in the simulations and at two successively doubled resolutions.  Figure~\ref{fig:thermostat_convergence} shows the cooling powers as a function of the timestep.  The primary cooling rate is essentially independent of temporal resolution, whereas the secondary cooling rate is halved whenever the timestep is halved and is four to five orders of magnitude weaker. Equivalently, the secondary cooling rate is tiny and scales linearly with $\Delta t$ and therefore vanishes in the continuum limit. The secondary thermostat thus only compensates for a numerical heating effect without modifying the physical thermostatting rate.}

 \end{document}